\documentclass[twocolumn,prb,aps,amsmath,amssymb,superscriptaddress,floatfix]{revtex4}

\usepackage{graphicx}
\usepackage{epstopdf}
\usepackage{epsfig}
\usepackage{amsbsy}
\usepackage{color}
\usepackage{mathtools}
\usepackage{cancel}

\DeclarePairedDelimiterX\braket[2]{\langle}{\rangle}{#1 \delimsize\vert #2}

\newcommand{\reff}[1]{(\ref{#1})}
\begin{document}
\title
{Spin transitions driven by electric dipole spin resonance  in fluorinated single- and bilayer-graphene quantum dots}

\author{D.P. \.Zebrowski}
 \affiliation{AGH University of Science and Technology, \\ Faculty of Physics and Applied Computer Science,
al. Mickiewicza 30, 30-059 Krak\'ow, Poland}

\author{F.M. Peeters}
 \affiliation{Departement Fysica, Universiteit Antwerpen,\\  Groenenborgerlaan 171, B-2020 Antwerpen, Belgium}

\author{B. Szafran}
 \affiliation{AGH University of Science and Technology, \\ Faculty of Physics and Applied Computer Science,
al. Mickiewicza 30, 30-059 Krak\'ow, Poland}

\date{\today}

\begin{abstract}
Spin transitions driven by a periodically varying electric potential in dilute fluorinated graphene quantum dots are investigated.
Flakes of monolayer graphene are considered  as well as  electrostatic electron traps
induced in bilayer graphene. The stationary states are obtained within the tight-binding approach and are used to the basis 
of eigenstates to describe  the system dynamics. The dilute fluorination of the top layer
lifts the valley degeneracy of the confined states and attenuates the orbital magnetic dipole moments due to  current circulation within the flake.
Moreover, the spin-orbit coupling introduced by the surface deformation of the top layer induced by the adatoms allows  spin flips to be 
driven by the AC electric field. For the bilayer quantum dots the spin flip times is substantially shorter than the experimental spin relaxation.
Dynamical effects including  many-photon and multilevel transitions are also discussed.
\end{abstract}

\maketitle

\section{Introduction}

Graphene \cite{novo} is a low defect material \cite{mfp}
with weak hyperfine interaction \cite{c13},
long spin coherence and relaxation times \cite{mass,weihan} 
 making it a promising medium for quantum information storage and processing based on the electron spins \cite{tr,mi,vo}.
However, the implementation of  spin quantum information circuitry  requires carrier confinement \cite{divi} which
in monolayer graphene is excluded by the Klein tunneling effect \cite{kt}. 
The energy gap required for such  electrostatic confinement can be induced  by lateral confinement in graphene nanoribbons \cite{tr,nrb1,nrb2,nrb3,nrb4,nrb5}.
Alternatively, a finite size flake of graphene \cite{flake1,flake2,flake3,flake4,flake5,flake6,mori,HexTriang,SingleExp} could be used as a quantum dot for the carrier confinement.
For both nanoribbons and nanoflakes the properties of the confined states are strongly influenced by the details of the edge which are hard to control.
The influence of the edges can be eliminated in a biased bilayer graphene which can host  quantum dots defined  electrostatically \cite{bi1,bi2,bi3,bi4,bi5,bi6,bi7,bi8}.

Spin-orbit coupling is a necessary prerequisite  for the manipulation of  confined spins by electric fields \cite{edsr1,edsr2,edsr3,edsr4,edsr5,edsr6,edsr7,edsr8,edsr9}.
In graphene the spin-orbit coupling is very weak since the intrinsic carbon atom  spin-orbit coupling 
is not transferred to the $\pi$ graphene band due to the orthogonality of $p_z$ orbitals with $p_x$ and $p_y$  forming the $\sigma$ bonds within the layer \cite{laird}.
In bilayer graphene the path for a stronger spin-orbit coupling could be opened by interlayer $\pi$ and $\sigma$ bands mixing \cite{gunjp},
which however turns out to be ineffective \cite{SoBilFabian}, and therefore resulting in a spin-orbit coupling energy exactly on the level of the single layer material.
However, the spin-orbit coupling can be introduced into graphene by adatoms \cite{ad}, e.g. hydrogen \cite{Hydro,Hydro1,Hydro2,Hydro3,Hydro4,Hydro5,Hydro6} or fluorine \cite{Fluory,BilayerFluorExp,Santos,BilayerFluor,Fluori1,Fluori2,Fluori3}. The fluorine adatoms, even in the dilute
limit \cite{Fluory},  increase the single-layer spin-orbit coupling energy substantially. A local deformation of the graphene sheet by flurorine produces $\sigma$-$\pi$ band mixing \cite{Fluory} as in folded graphene or carbon nanotubes \cite{laird,be1,be2,be3,be4,be5,be6} givings rise to  spin-dependent hopping near the adatom.

The purpose of the present paper is the characterization of the spin-flip electric-dipole spin resonance (EDSR) \cite{edsr1,edsr2,edsr3,edsr4,edsr5,edsr6,edsr7,edsr8,edsr9} in a bilayer graphene quantum dot by external nano-engineered electrostatic potential, in which the spin-orbit interaction is introduced by fluorine adatoms at the top layer.
We consider dilute adatom concentrations, for which the band structure of graphene is perturbed to a minimal extent. 
We consider the system near the neutrality point and determine the Hamiltonian eigenstates using the atomistic tight binding approach
that are next used for description of the spin resonance driven by AC electric field. 
We find that already at 0.5\% concentration of the fluorine adatoms the spin-flip transition time driven by the AC field
 are shorter than the spin-relaxation times \cite{weihan}, which should allow for an experimental induction and detection of spin-flips in Pauli-blocked
quantum dots \cite{edsr1,edsr2,edsr3,edsr4,edsr5,edsr6,edsr7,edsr8,edsr9}.
\begin{figure}[htbp]
 a)\hspace{-0.0cm}\includegraphics[scale=0.06]{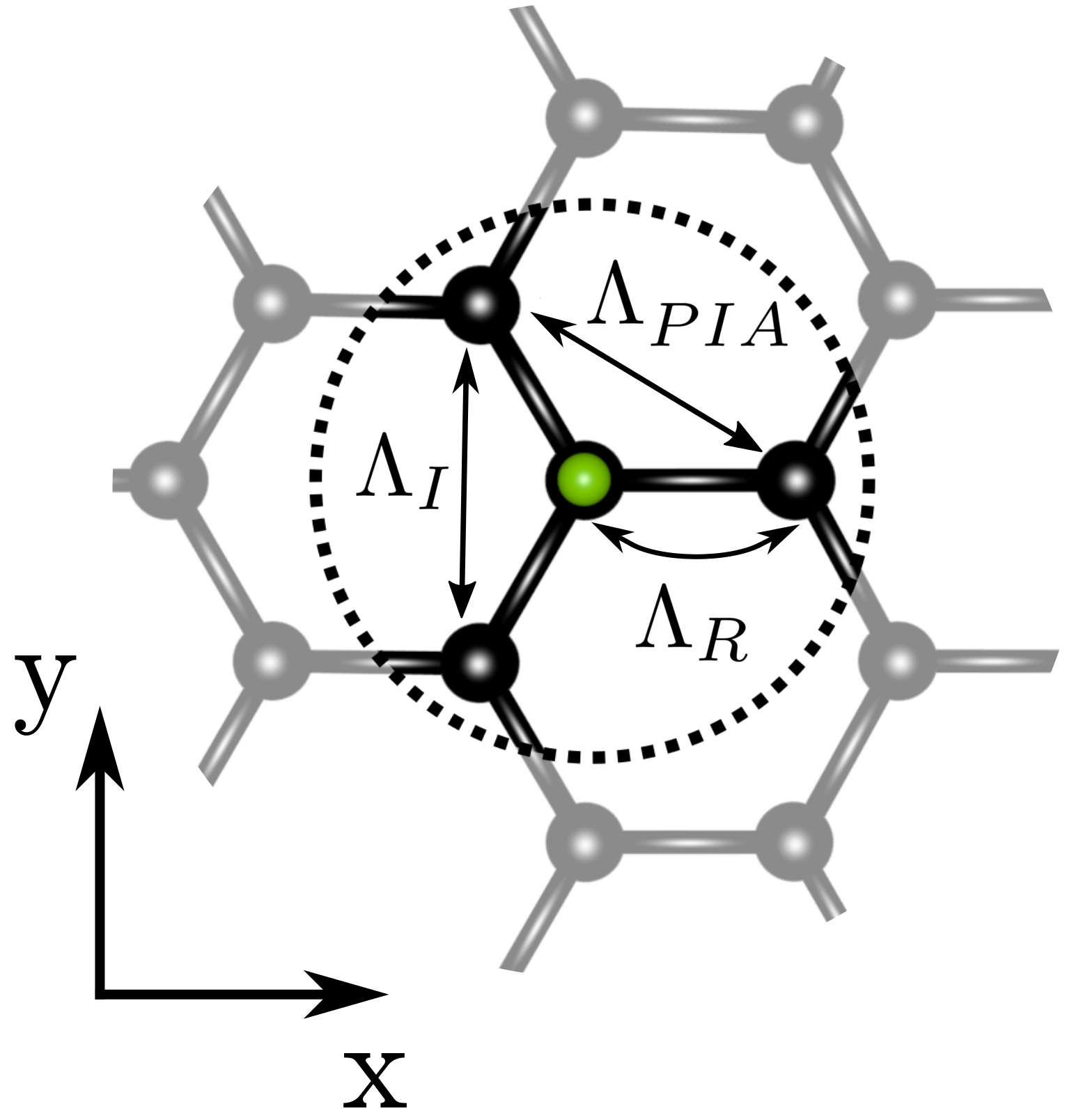}\hspace{0.3cm}b)\hspace{-0.53cm}\includegraphics[scale=0.22]{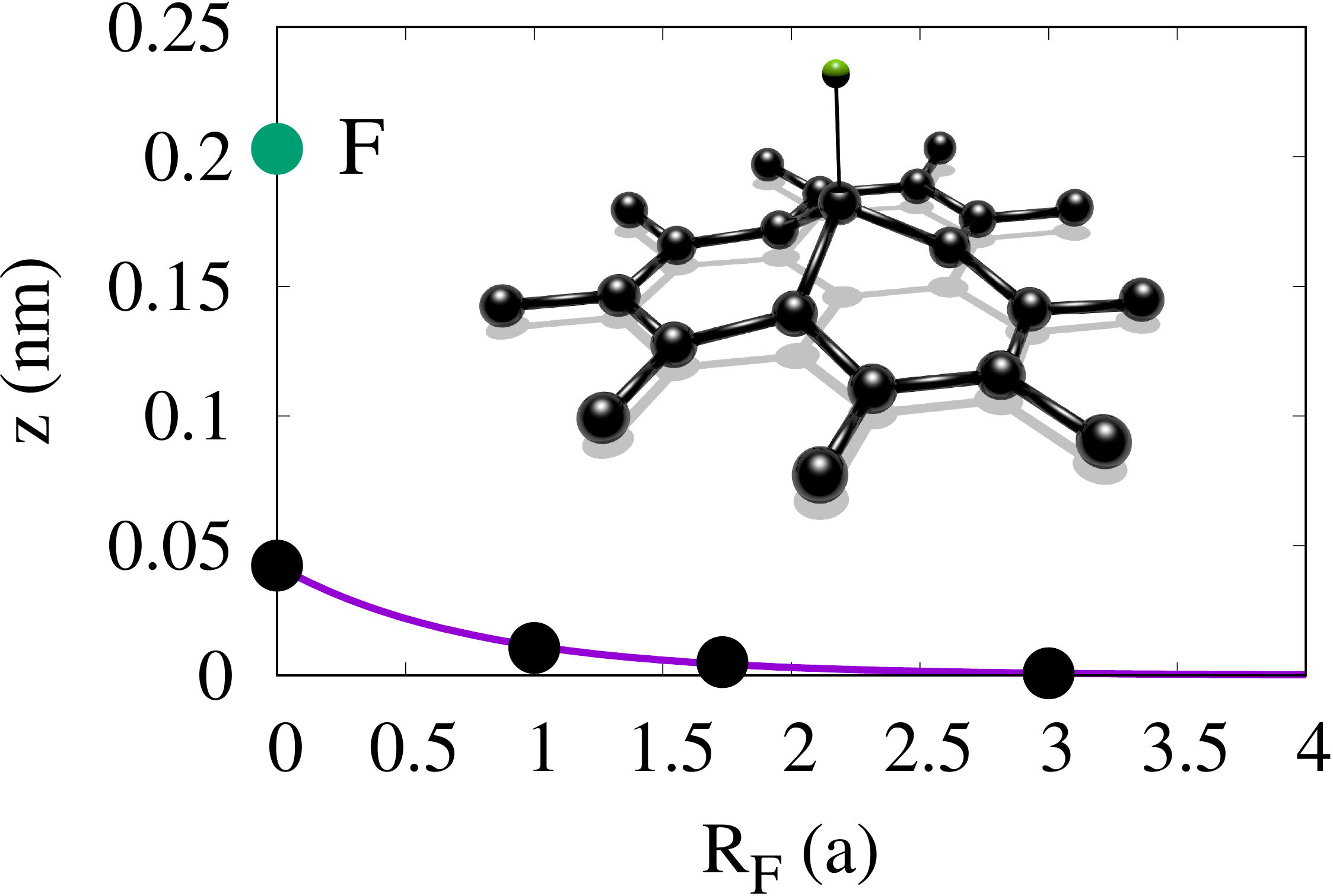} 
\label{Rys1}
\caption{(a) Top view of the monolayer carbon (black) lattice with the additional fluorine atom (green) above.
The $\Lambda$'s correspond to spin-orbit coupling terms introduced by fluorine adatom. 
 The  dashed circle shows the area where spin-orbit terms induced by fluorine adatom are non zero. 
(b) The position of the fluorine and the perpendicular displacement of the carbon atoms induced by the adatom - after Ref. \cite{BilayerFluor}
in units of $a=1.42$ \AA\; - the nearest-neighbor carbon atoms distance. The line $z/a=\alpha\exp(-\beta R_F/a)$ is a fit to the data 
, where $R_F$ is the distance to the fluorinated carbon atom, $\alpha=0.0423$ and $\beta=1.316$. 
{ The inset shows a 3-D view of the fluorinated carbon layer.} }
\label{ukladf}
\end{figure}
\section{Theory}
\subsection{Monolayer Hamiltonian}

We use the tight-binding Hamiltonian for fluorinated graphene following Ref. \cite{Fluory} in the dilute limit. The Hamiltonian reproduces the first principles
electron structure near the Dirac point within the basis of $p_z$ orbitals -- for the carbon and the fluorine atoms, only. The Hamiltonian has the general form
\begin{equation}
  \hat{H}=\sum_{i,\sigma}\epsilon^{\sigma}_{i,\sigma}\hat{c}^{\dagger}_{i,\sigma}\hat{c}_{i,\sigma}+\sum_{ ij,\sigma\sigma'} \left( \hat{c}^{\dagger}_{i,\sigma}t^{\sigma\sigma'}_{ij}\hat{c}_{j,\sigma'}+h.c \right ),
  \label{Hfull}
\end{equation}
where $\hat{c}^{\dagger}_{i,\sigma} $($\hat{c}_{i,\sigma}$) is the creation (annihilation) operator at ion $i$ with spin $\sigma$. 
The $\epsilon^{\sigma}_i$ and $t^{\sigma\sigma'}_{ij}$ are the on-site and the hopping energies respectively. The zero of the energy is taken such that the on-site energies $\epsilon^{\sigma}_i$ for monolayer graphene are equal to $0$ for the 
carbon atoms and $\epsilon_F=-2.2$ eV for the fluorine site. The spin-conserving hopping energy between nearest neighbour carbon atoms are equal to $t_{ij}=-2.6$ eV and 
$t^{\sigma\sigma}_{ij}=5.5$ eV for the carbon-fluorine bond. 

The adatom  introduces spin-orbit coupling terms by the hopping parameters for the bonds
 in the vicinity [see Fig. 1(a)] of the fluorinated carbon atoms \cite{Fluory}. 
The coupling is given \cite{Fluory} as a sum of several terms: 
(i) The intrinsic spin-orbit is locally enhanced 
\begin{equation}
 t^{\sigma\sigma'}_{ij}=i\frac{\Lambda_I}{3\sqrt{3}} \nu_{ij}\hat{S_z}, \label{int}
\end{equation}
with a largely enhanced coupling constant $\Lambda_I=3.3$ meV. The term 
mediates the spin conserving hopping between the next nearest neighbor carbon atoms \cite{KaneMele},
which are neighbours of the fluorinated carbon [see Fig. 1(a)]. 
The factor $\nu_{ij}$ takes the value +1 (-1) for the hopping path between $j$ and $i$ nodes passing through a common neighbour $k$ with a counterclockwise (clockwise) turn. 

(ii) Another term is called pseudo-inversion-asymmetry \cite{Fluory} (PIA) which mediates the spin-flip hoping between the nearest neighbours of the fluorinated carbon atom 
\begin{equation}
 t^{\sigma\sigma'}_{ij}=2i\frac{\Lambda_{PIA}}{3} \hat{\vec{S}}\times \vec{n}_{ji}, 
\end{equation}
where  $\Lambda_{PIA}=7.3$ meV, $\hat{\vec{S}} $ is the vector of spin component operators and $ \vec{n}_{ij}$ is a unit vector that points from node $j$ to $i$.

(iii) The last term  introduces the spin-orbit coupling between the fluorinated carbon atom and its three nearest neighbours.
\begin{equation}
 t^{\sigma\sigma'}_{ij}=2i\frac{\Lambda_R}{3}\hat{\vec{S}}\times \vec{n}_{ij}, \label{Ras}
\end{equation}
with $\Lambda_R=11.2$ meV. This term is of the Rashba-type 
and results from the local electric field induced by the adatom.

Besides the spin-orbit coupling hoppings introduced by the adatoms, 
the tight-binding Hamiltonian \cite{SoFabian2,SoBilFabian} contains 
intrinsic and  Rashba  spin-orbit coupling terms in the form given by Eqs. \eqref{int}
and \eqref{int}, which couple all the next-nearest neighbors and the nearest neighbours within the layer, respectively.
The intrinsic coupling constant $\lambda_i$ that replaces $\Lambda_I$ in Eq.   \eqref{int} is  only $\lambda_i=12 \mu eV $.

 In order to induce the energy gap for confinement in the bilayer graphene we 
consider   electric fields of the order of 1V/nm. This corresponds to an electrostatic
potential difference of about 0.3 eV \cite{kanaya,BilayerField1Vnm,santoss} within the layers spaced by 3.32 \AA\;. For 
the electric field of 1V/nm  the Rashba spin-orbit coupling constant that replaces $\Lambda_R$ in Eq. \eqref{Ras} is $\lambda_R=5\mu$eV \cite{SoFabian2}. 
Already at 0.5\% concentraction of the fluorinated carbon atoms, the intrinsic and Rashba spin-orbit interactions characteristic for clean graphene 
produce negligible corrections to the energies of confined states and spin-flip times.

The external perpendicular magnetic field is taken into account by including the Peierls phase
in the hopping parameters
\begin{equation}
t^{\sigma\sigma'}_{ij} \rightarrow t^{\sigma\sigma'}_{ij}e^{i\phi_{ij}}, 
\end{equation}
with   $\phi_{ij}=\frac{e}{\hbar}\int_{\mathbf{r_i}}^{\mathbf{r_j}}\mathbf{A}\cdot \mathbf{dl}$ and $\mathbf{B}=\mathbf{\nabla} \times \mathbf{A}$. 
Moreover, the on-site spin-dependent energies must now include the external magnetic field induced  spin Zeeman splitting
\begin{equation}
\epsilon^{\sigma}_{i,\sigma}\rightarrow \epsilon^{\sigma}_{i,\sigma}+\frac{1}{2}\mu_Bg\sigma^zB_z,
\end{equation}
where $g=2$ is the Land\`e factor.
\subsection{Bilayer graphene}

\begin{figure}[htbp]
\includegraphics[scale=0.15]{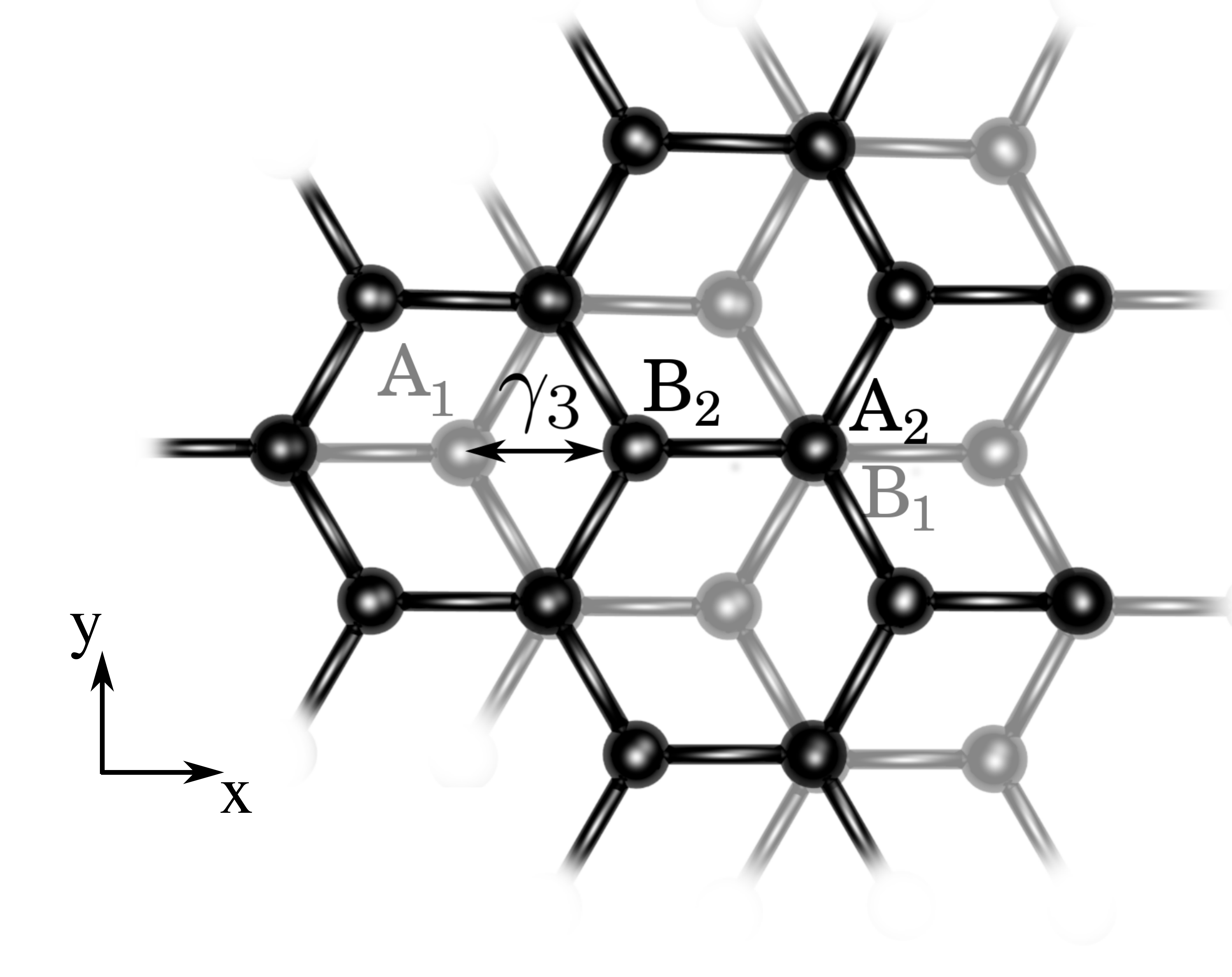}
\caption{Top view of  bilayer graphene in Bernal stacking. The vertical dimers link the atoms of the upper sublattice A (A2) with the 
sublattice B of the lower layer (B1). The vertical hopping energy  along the dimer are included in the calculations
as well as the nearest neighbor direct interlayer hopping -- labelled by $\gamma_3$ in the figure.  } \label{sze}
\end{figure}

We consider graphene bilayers \cite{bi1} in Bernal stacking (Fig. \ref{sze}) and the fluorine atoms adsorbed by the upper layer \cite{Santos}.
We assume that the spin-orbit coupling interactions induced by the fluorine adatoms are limited to the nearest neighbours of the fluorinated carbon atom in the upper layer \cite{SoBilFabian},
which is justified by the weaker coupling along the vertical interlayer van der Waals bond and the large ratio of the distance of the nearest  interlayer and intralayer  neighbors \cite{bi1}
$a_{A1,B2}=3.32$ \AA\; and $a_{A2,B1}=1.42$ \AA, respectively.

We account for the deformation of the top graphene plane \cite{Fluory,BilayerFluor,Santos} 
-- see Fig. \ref{ukladf}(b) -- and assume that  the lower graphene plane is flat. In that way the modification of the distance between the layers in the present model is an upper bound to the actual case. 
The variation of the interlayer distance is accounted for by the Harrison \cite{Harrison} $d^{-2}$ law. 
Accordingly, the interlayer hopping parameters are scaled  by a factor of $(d_{ij}/d_{ij}^F)^2$, where $d_{ij}$ ($d_{ij}^F$ ) is the distance between the ions without (with) the fluorine adsorption. 
For the interlayer hopping we take the nearest-neighbor vertical one that goes along the (A2-B1) dimer (see Fig. \ref{sze}), with $t_{ij}=-0.3$ eV in the absence of adatoms.
We also account for the skew interlayer nearest-neighbor direct hopping between the carbon atoms
that do not form dimers [Fig. \ref{sze}]. This coupling produces trigonal warping \cite{tri} of the Fermi level near the Dirac points. 
Since we deal with localized states a contribution of the wave vectors off the Dirac points is likely,
hence the inclusion of skew hoppings within the model. We take $\gamma_3=-0.2$ eV for the value of the hopping that is within the range of usually applied parameters that varies betwen -0.1 and -0.38 eV \cite{bi1}. 
No spin-dependent hopping between the layers is introduced according to the conclusion of Ref. \cite{SoBilFabian}.

The Harrison law rescales the vertical and skew hopping parameters by a factor of 0.79 and by a factor of 0.97, respectively.   
Although the present modification of the interlayer distance is an upper bound to the actual one, 
the corrections of the Harrison law to the hopping elements as well as the ones due to the  skew hopping elements  have  only a very weak effect on the energies of the confined states
 in the considered dilute limit of fluorine adatoms.  Nevertheless, the skew hopping  influences  the transition rates (see below).

\subsection{ The confinement potential}
\begin{figure}[htbp]
\includegraphics[scale=0.3]{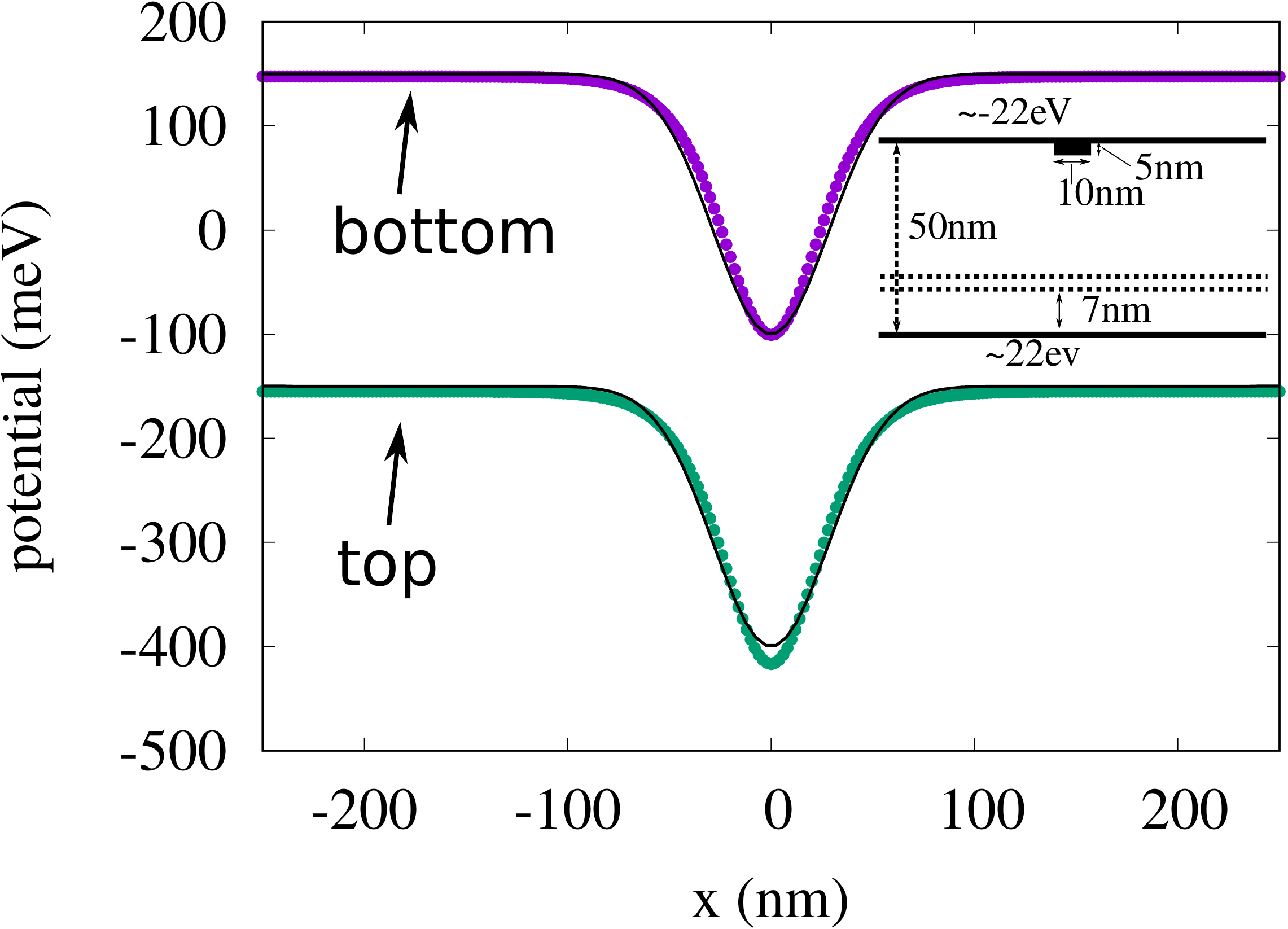}
\caption{
Inset: our model structure:  bilayer (dashed lines) graphene suspended between the metal gates. 
The top gate has a cylindrical protrusion that forms the lateral confinement in the layers.
In the main figure the dotted curve is the solution of the Laplace equation and the coloured lines a  Gaussian fit.
} 
\label{ukladsz}
\end{figure}

We consider the bilayer graphene near the neutrality point. For the choice of the model potential we solved the Laplace equation for bilayer graphene suspended between two metal gates [Fig. \ref{ukladsz}] which induce the electric field  perpendicular to the graphene layers. The top gate has a 
protrusion in its center [inset to Fig. \ref{ukladsz}] that should form a lateral component of the electric field and form 
the confinement potential within the system. An attractive potential is applied to the top gate. The attraction by the protrusion should be stronger and
form the quantum dot within the layer (alternatively an opening in the split gate with a repulsive potential could be used).
 The electrostatic potential for the structure depicted in the inset of Fig. \ref{ukladsz} was solved using the Laplace equation for a constant potential at the metal and a vanishing normal electric field at the sides of the structure. 
 Finite difference approach was applied.  For  potentials of $\pm$ 22 V applied to the gates the confinement potential at the top and bottom layers is given by the dotted curves in Fig. \ref{ukladsz}, which can be fitted  
by a Gaussian parametrization of the potential energy $V=-V_0 \exp(-r^2/r_0^2)\pm \frac{W_{bias}}{2}$ with $V_0=250$ meV and $W_{bias}=300$ meV, and $r_0$ defines the range of confinement.
Notice that the difference of the electrostatic potential between the layers is nearly constant inside and outside of the potential cavity.
Based on this finding, we consider the following model potential for bilayer graphene  
\begin{equation}
\begin{split}
W(x,y,z) =&\left(-0.5+\frac{z}{a_i}\right) W_{bias} \\
&-W_{QD} \exp\left(- \frac{x^2+y^2}{r_0^2} \right),
\end{split}
\end{equation}  where $a_i=3.32$ \AA\; is the interlayer distance.

For the atomistic approach that we use in the following  we take a hexagonal flake of side length of 36.4 nm. 
{ We set the radius of the potential to be  $r_0=4$ nm and thus  the confined electrons will not reach  the edge of the flake.}


%
\subsection{Dynamics}

We study spin transitions driven by an external AC electric field \cite{edsr1,edsr2,edsr3,edsr4,edsr5,edsr6,edsr7,edsr8,edsr9} using the numerical
solution to the time-dependent Schr\"odinger equation. A single excess electron
confined within the quantum dot induced in the bilayer graphene flake or a single excess electron in a monolayer flake is investigated. 
The excess electron occupies the low-energy part of the spectrum within the conduction band that is separated from the valence band by 
an energy gap open by the quantum size effect (monolayer) and / or by the bias field between the layers (bilayer). 
In systems for which the time-dependent simulations are performed the energy gap is of the order of 150 meV or larger. For that reason we consider
all the states of the valence band frozen and occupied by electrons. The driven electron
states are then calculated  in the basis of the conduction band states only,
 
We solve the time-dependent Schr\"odinger equation
\begin{equation}
 i\hbar\frac{\partial\Psi}{\partial t}=\hat{H}'(t) \Psi,
 \label{timeH}
\end{equation}
with the time-dependent Hamiltonian to include effects of the AC electric field applied within the graphene plane
\begin{equation}
 \hat{H}'(t)=\hat{H}+eFx\sin(\omega t),
\end{equation}
where $F$ and $\omega$ are the amplitude and frequency of the AC field, respectively.
We solve this equation in the basis spanned by the conduction band eigenstates of the Hamiltonian \reff{Hfull}
\begin{equation}
 \Psi(\mathbf{r},\sigma,t)=\sum_n c_{n}(t)\Psi_n(\mathbf{r},\sigma)e^{-i\frac{E_n t}{\hbar}}. \label{baa}
\end{equation}

After plugging this form into Eq. \reff{timeH} we obtain a set of differential equation for the coeffiecents $c_n(t)$
\begin{equation}
 i\hbar c'_k(t)=\sum_nc_n(t)eF\sin(\omega t)\langle \Psi_k |x|\Psi_n \rangle e^{-it\frac{E_n-E_k}{\hbar}}. 
\end{equation}
We solve these equations using the explicit Askar-Cakmak scheme\cite{Askar}.
We take up to 80 eigenstates into the basis \eqref{baa}. 
For monolayer flake we account for the lowest-energy states for the conduction band,
and for the bilayer quantum dots we also account for the dot-localized states 
with  energies within the energy gap.
The approach  gives a numerically exact solution of the 
system dynamics taken into account  non-linear and multilevel 
effects far outside the first-order two-level Rabi transitions. 

\begin{figure*}[htbp]
\begin{tabular}{c@{\hskip 0.1cm} c @{\hskip 0.1cm} c c} 
 \includegraphics[scale=0.25]{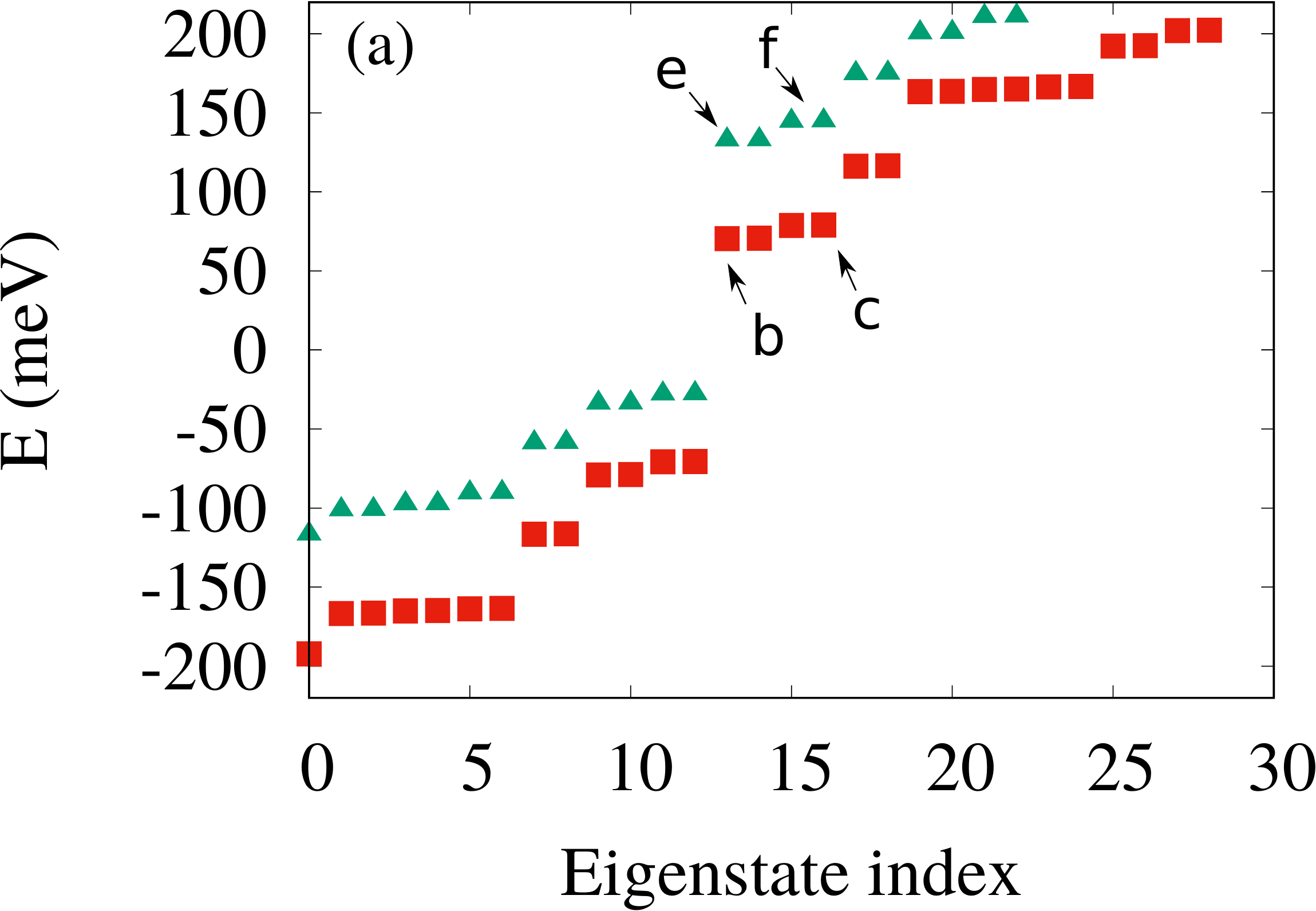} &  \includegraphics[scale=0.25]{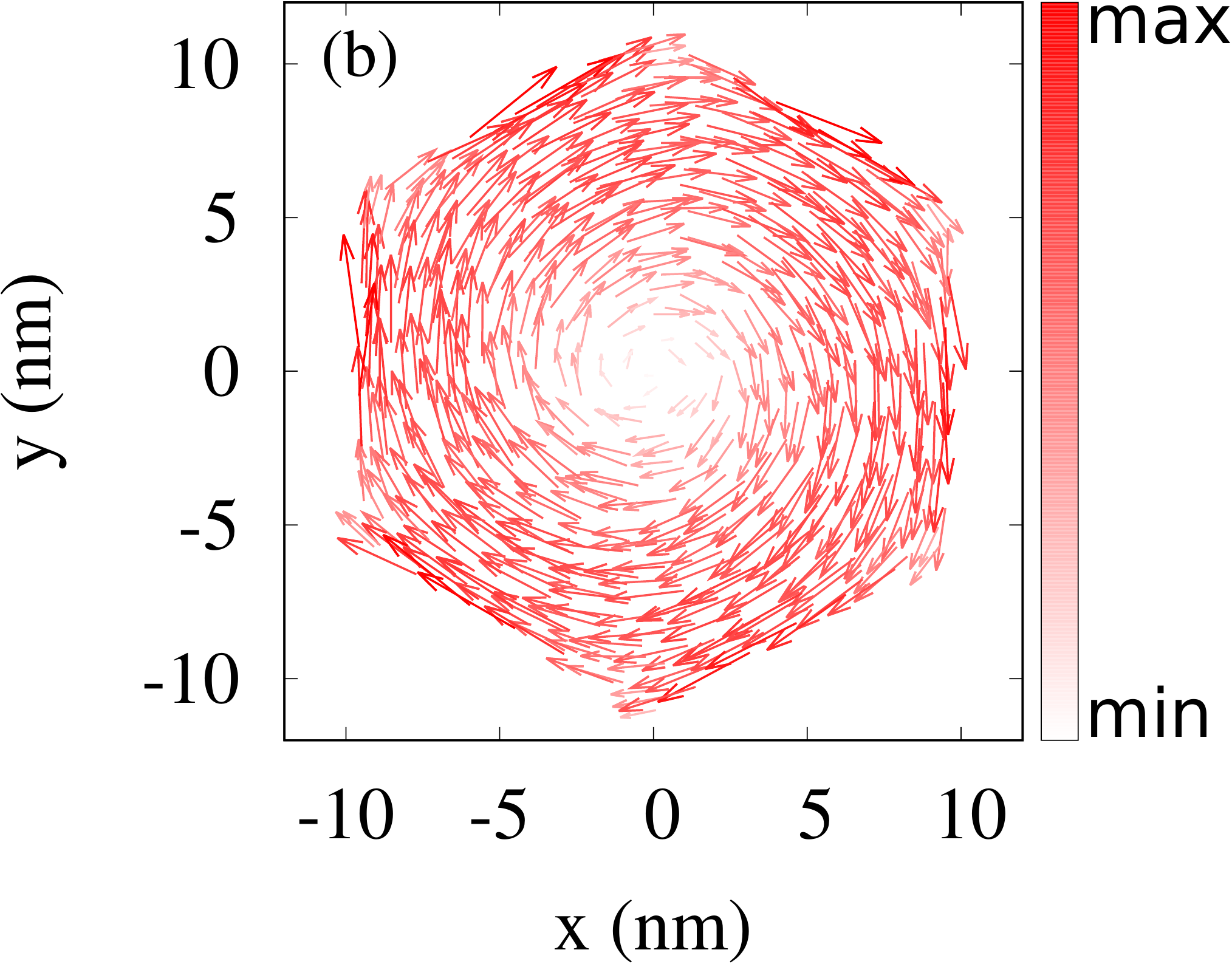} & \includegraphics[scale=0.25]{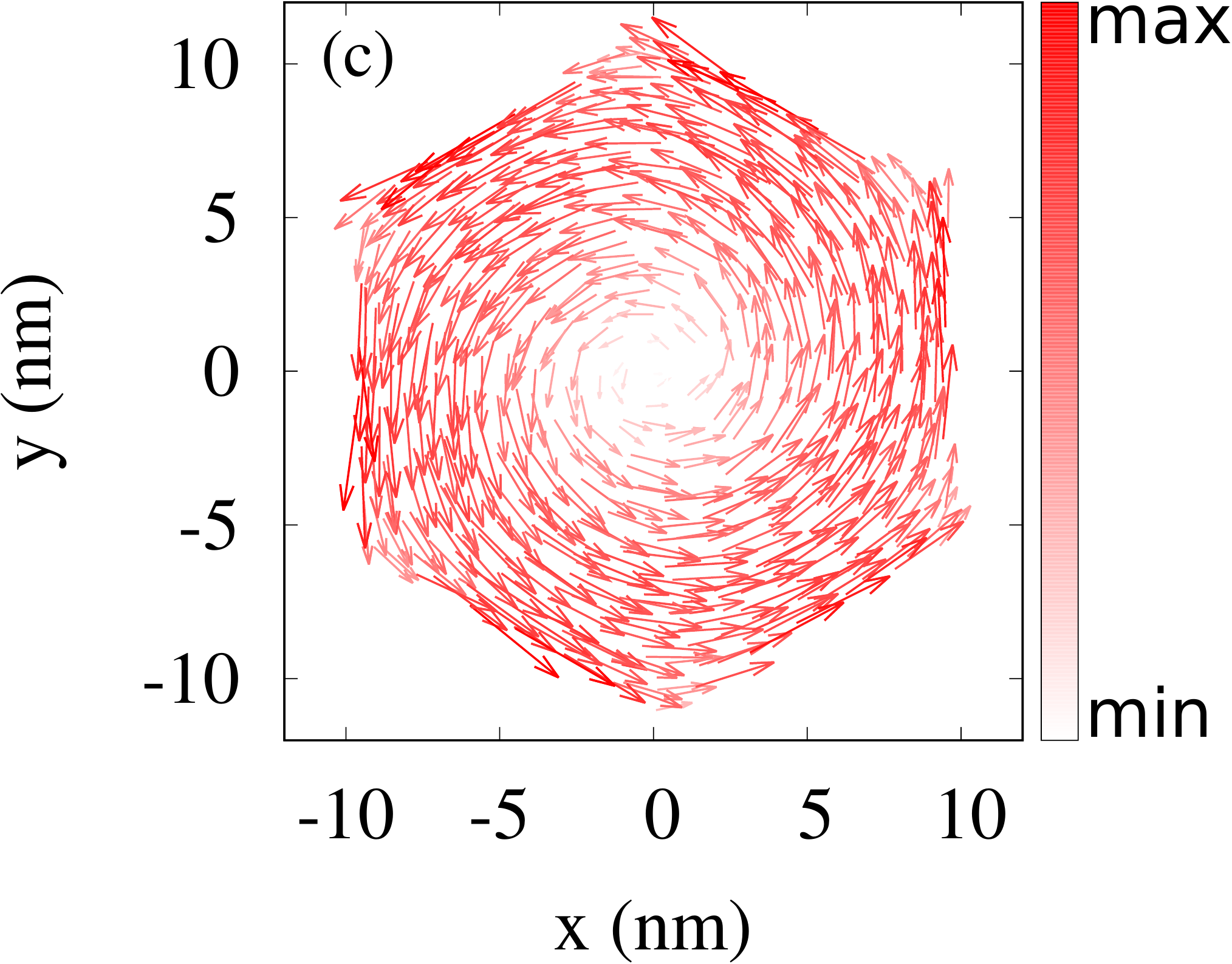} \\

\includegraphics[scale=0.25]{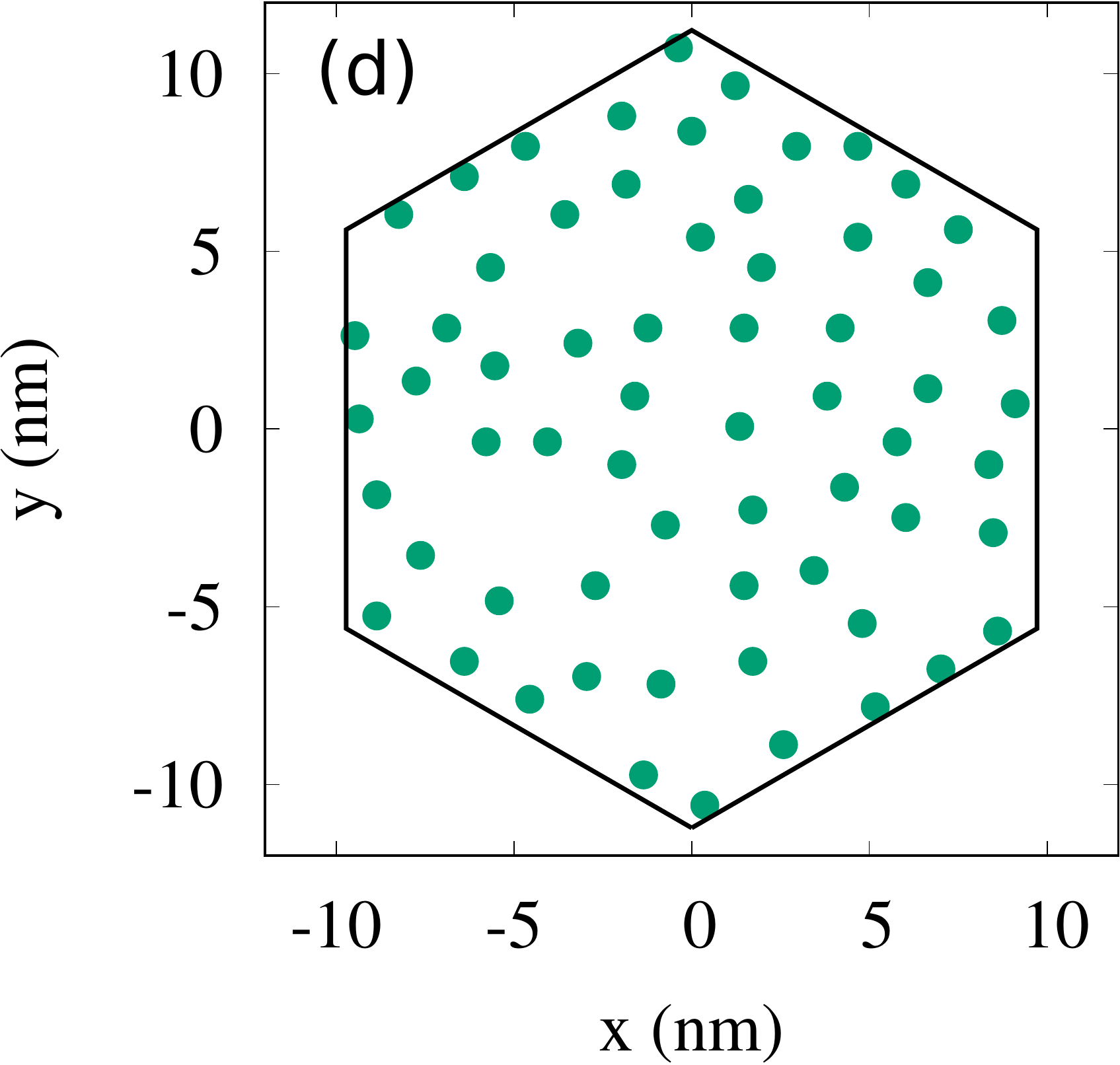} &  \includegraphics[scale=0.25]{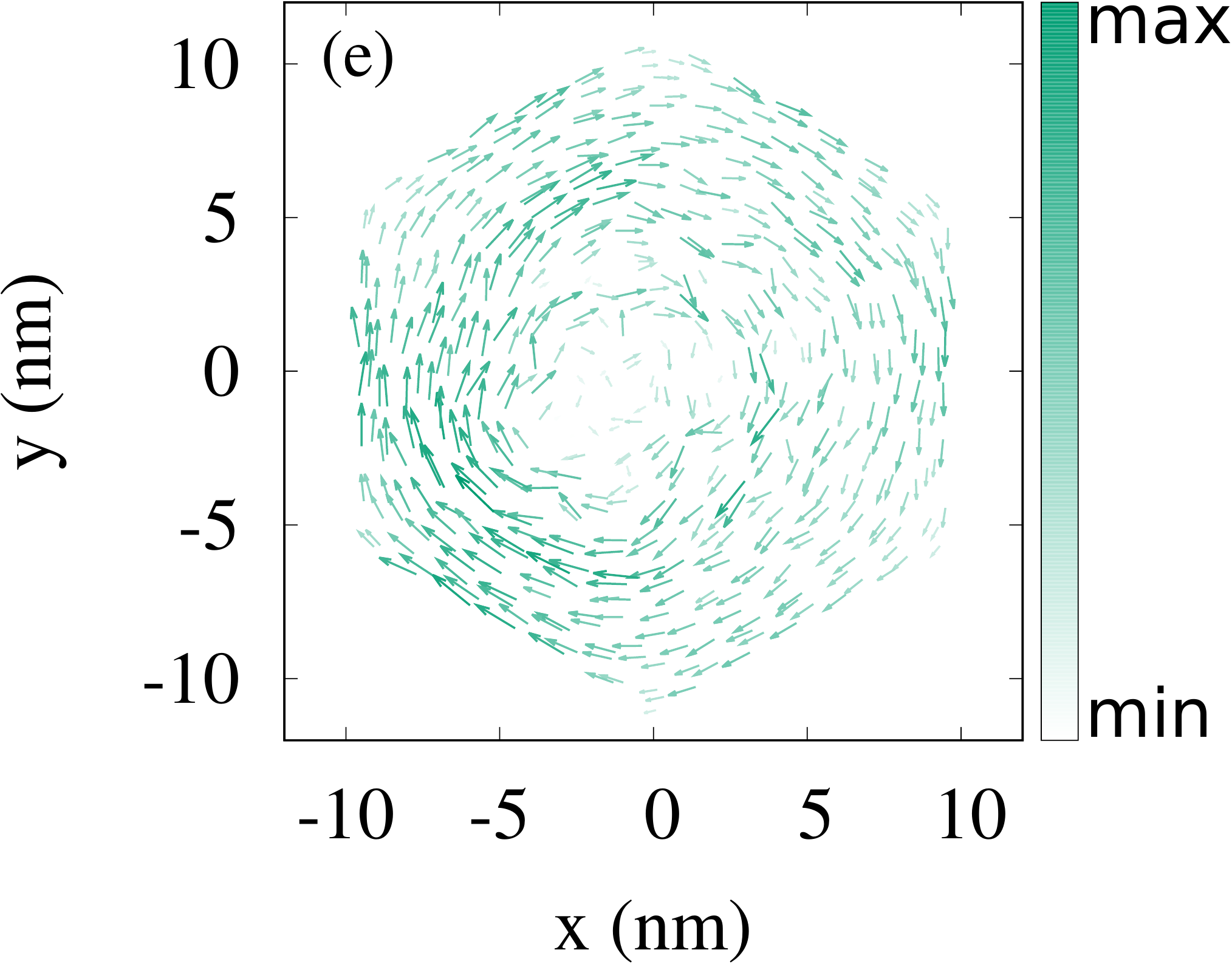} &  \includegraphics[scale=0.25]{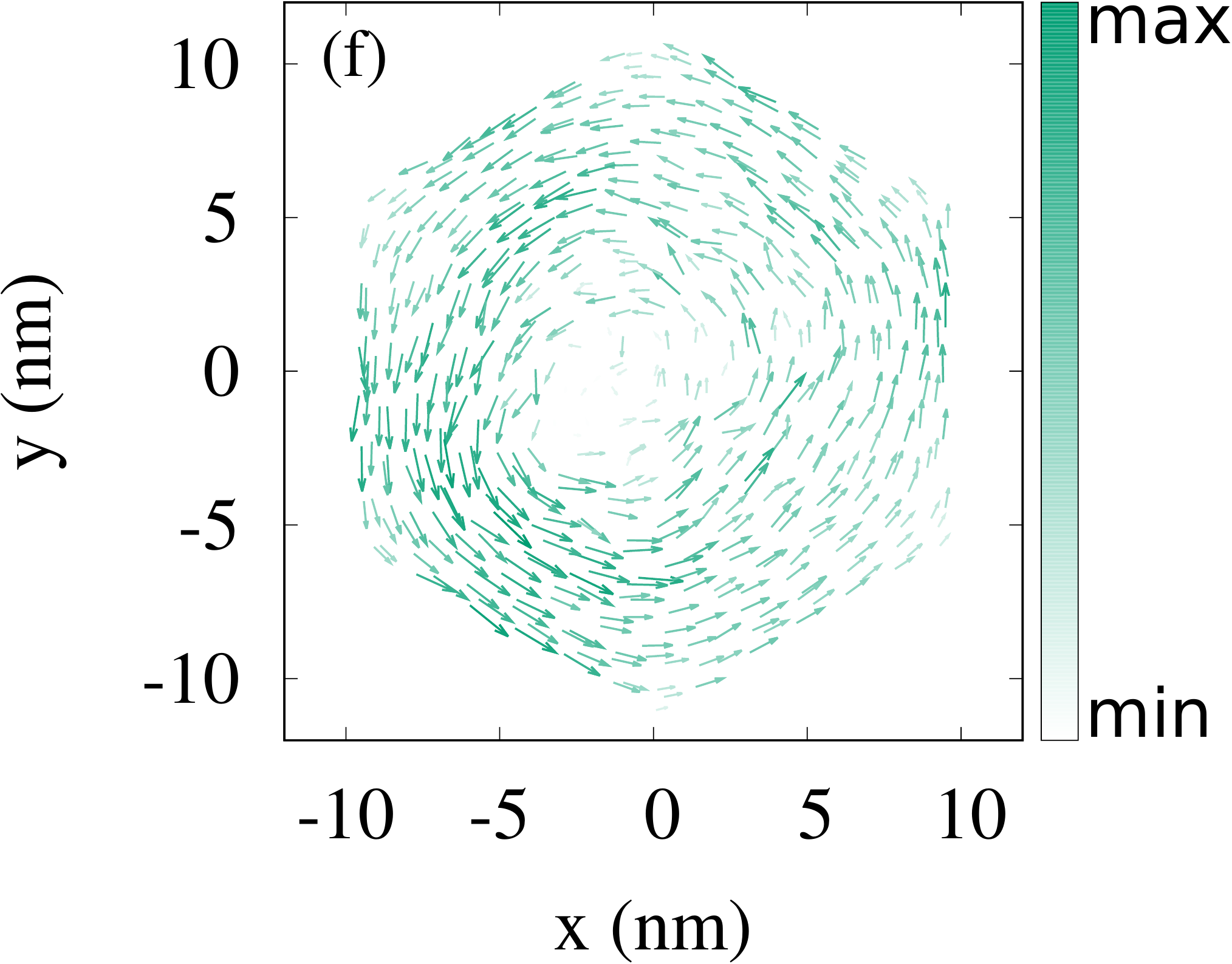}    \\ 
\end{tabular}

  \caption{(a) Energy spectrum as a function of eigenvalue index for a hexagonal flake. The red squares (green triangles) correspond to the case without (with) the fluorine adatoms. (b),(c),(e) and (f) 
  are current density profiles for the states indicated in (a). (d) The distribution of the fluorine adatoms on the flake. }

\label{flake}
\end{figure*}
\begin{figure}[htbp]
\includegraphics[scale=0.25]{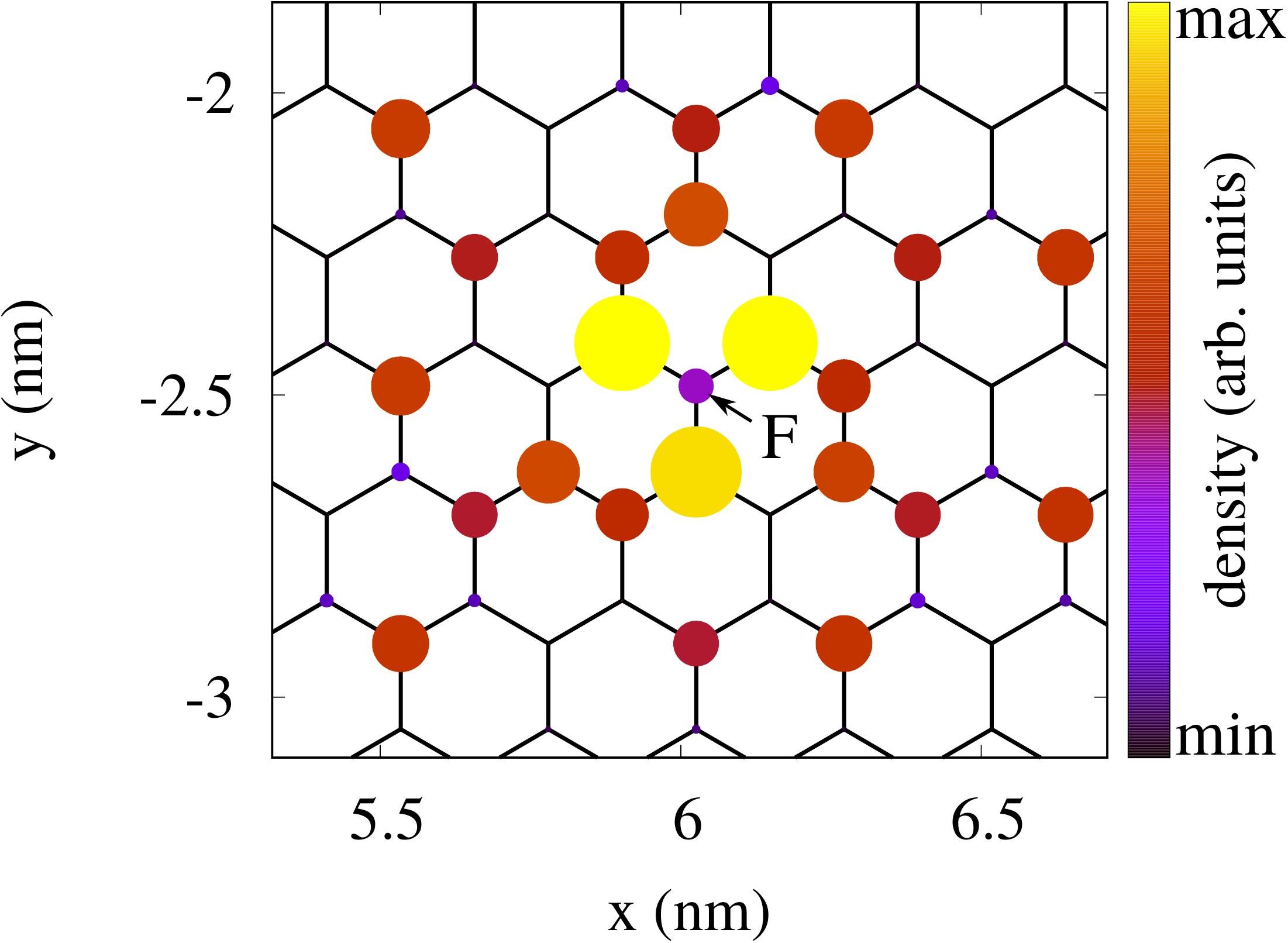}
  \caption{Electron density for the eigenstate marked by (e) in Fig. \ref{flake} near one of the fluorinated carbon atoms. The radius and the color
of the circle are proportional to the electron occupation for a given ion.
The density at $F$ shows the electron on the fluorinated carbon atom. }

\label{edens}
\end{figure}
\begin{figure}[htbp]
	
\includegraphics[scale=0.31]{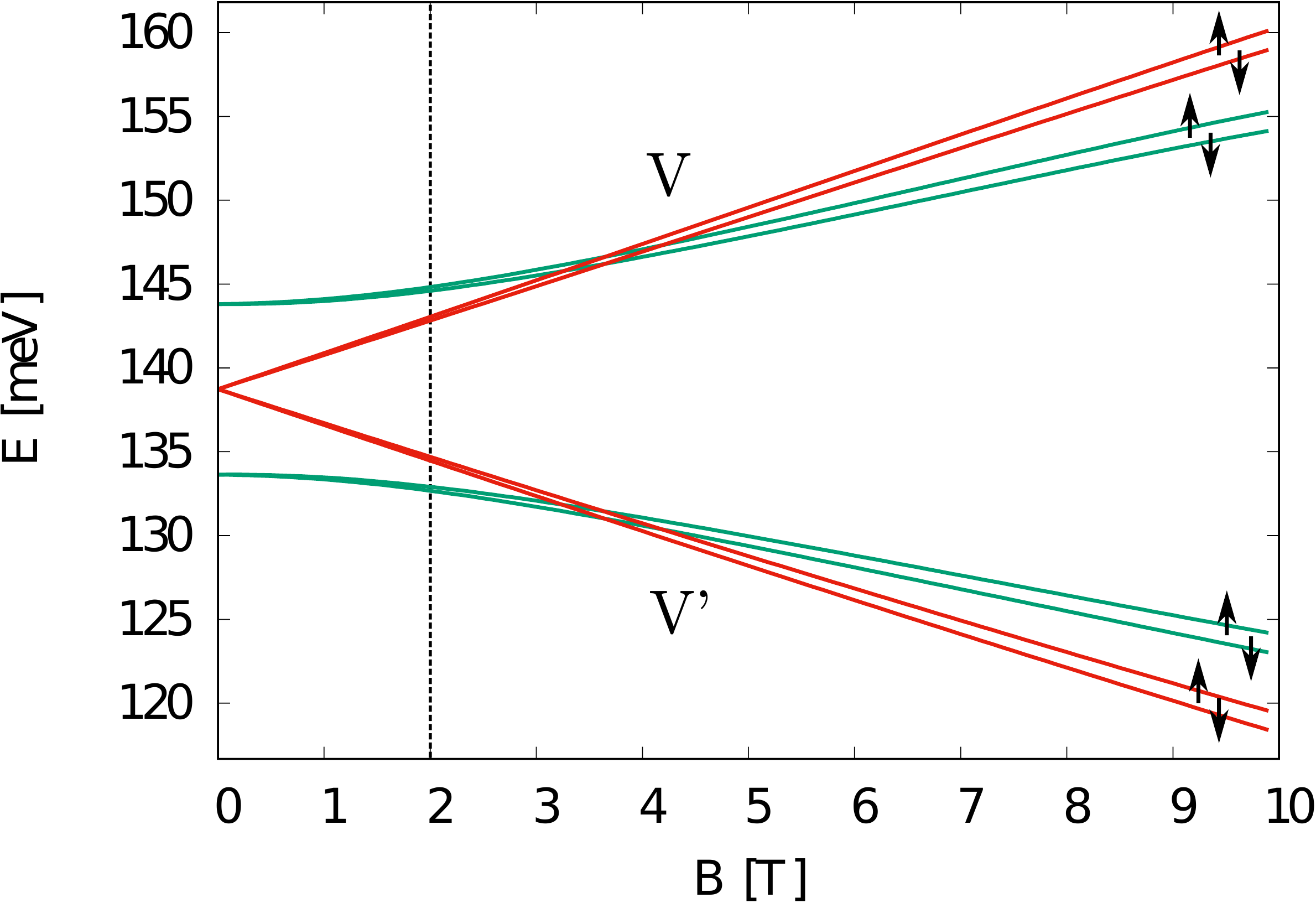}
 \caption{The four lowest-energy conduction-band levels of the monolayer grapehene flake  quantum dot as a function of applied magnetic field. The green (red) lines correspond to a flake with (without) fluorine adatoms.  The red lines are shifted up by $64$ meV for clarity. The vertical dashed line marks the field of 2T which was applied in Fig. \ref{flake} and which is set for the driven spin-valley transitions  in the following.  The arrows indicate the direction of the spin and the $V$, $V'$ label the valley states that increase and decrease in the external magnetic field, respectively.}\label{flakee}
\end{figure}
\begin{figure*}	
\begin{tabular}{c@{\hskip 1cm} c}
\hspace{-0.3cm}\includegraphics[scale=0.35]{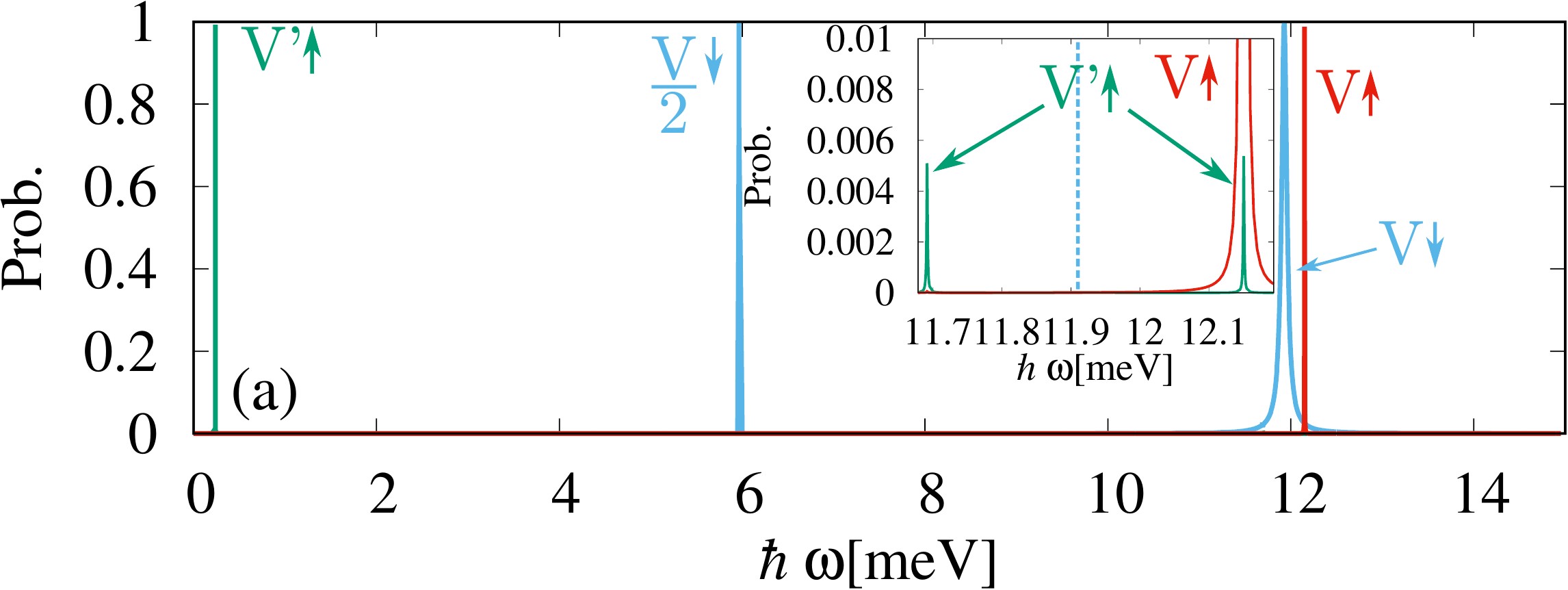} &  \hspace{-0.3cm}\includegraphics[scale=0.35]{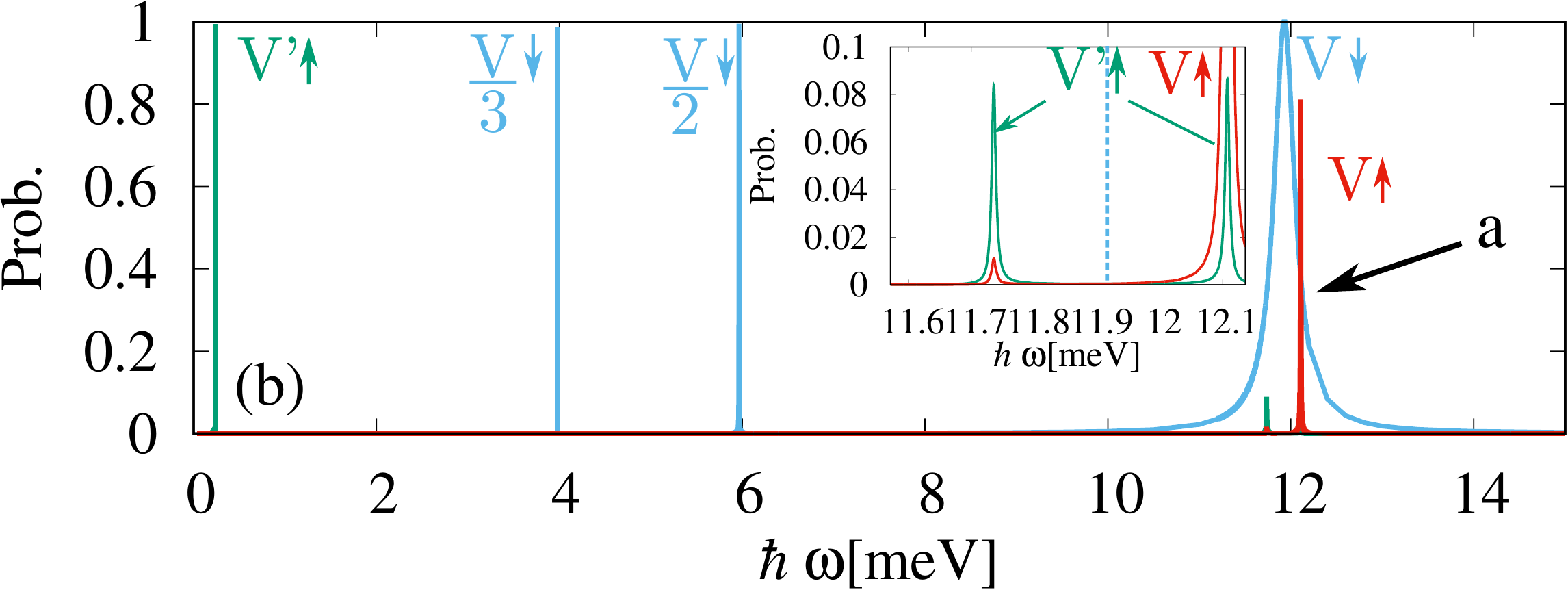} \\

\hspace{-0.3cm}\includegraphics[scale=0.35]{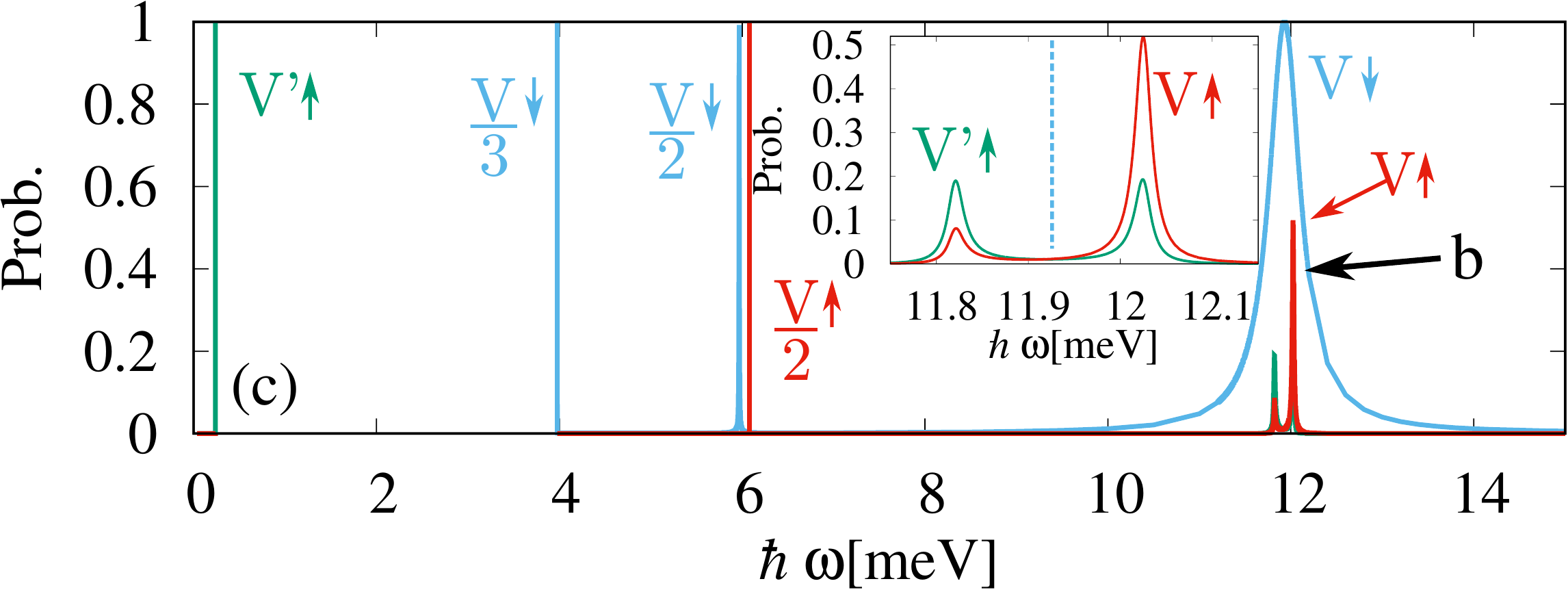} & \hspace{-0.3cm}\includegraphics[scale=0.35]{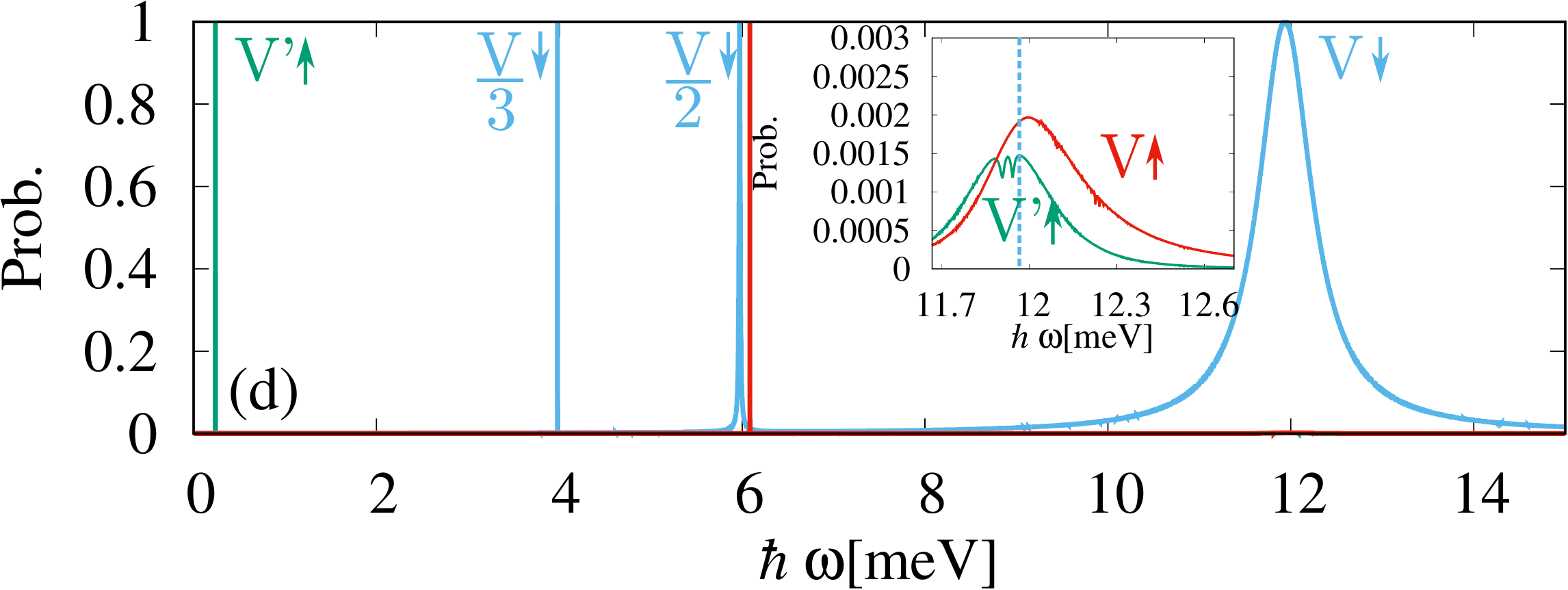} \\

\end{tabular}

\caption{
Maximal occupation probability of the three excited  states of the quadruple of Fig. \ref{flakee} for the fluorinated graphene flake at $B_z=2$ T as function
of the driving electric field frequency. The results were obtained from time-dependent simulations within a time span of 2 $\mu$s
and $V'\downarrow$ ground state is taken for the initial condition.  The amplitude of the in-plane AC electric field 
if $F=0.2$ kV/cm in (a), 0.8 kV/cm in (b), 1.2 kV/cm in (c) and 2 kV/cm in (d). The insets in (b-d) show the zoom for the spin-flipping transitions near $\hbar\omega_{AC}=12$ meV.
}\label{Trans2}
\end{figure*}
\begin{figure}[hbt]
\hspace{-0.3cm}\includegraphics[scale=0.35]{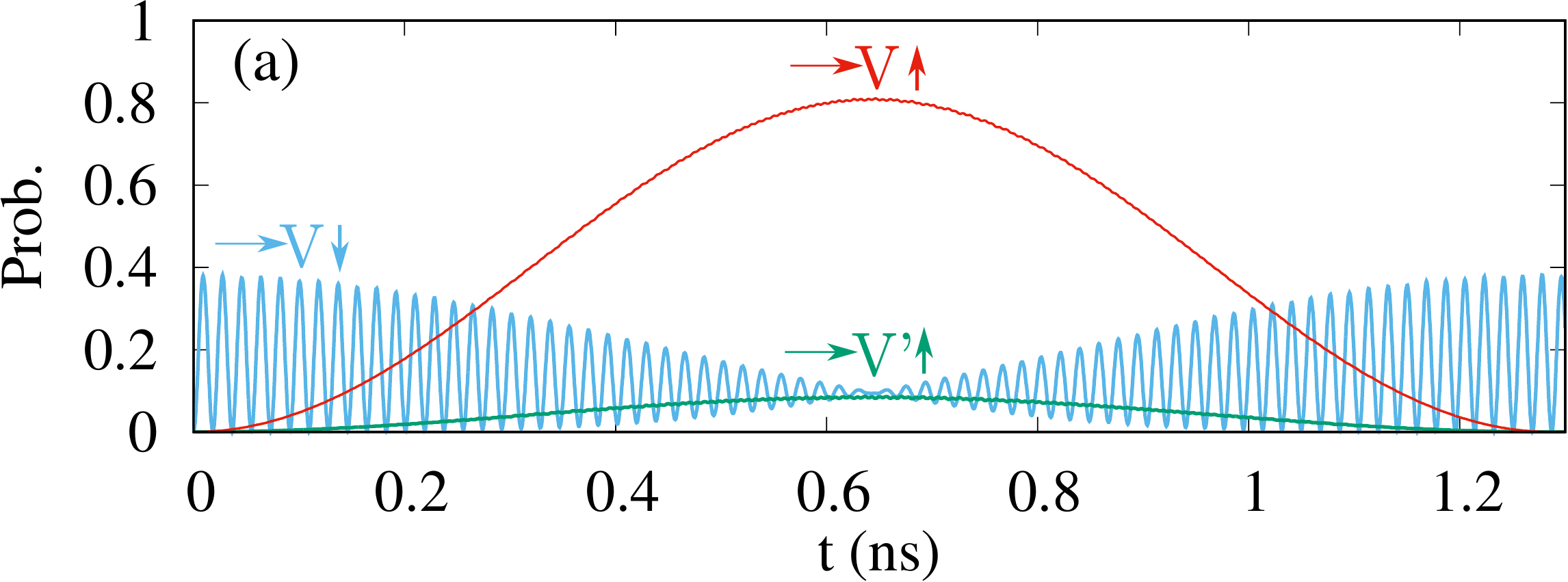} \\
\hspace{-0.3cm}\includegraphics[scale=0.35]{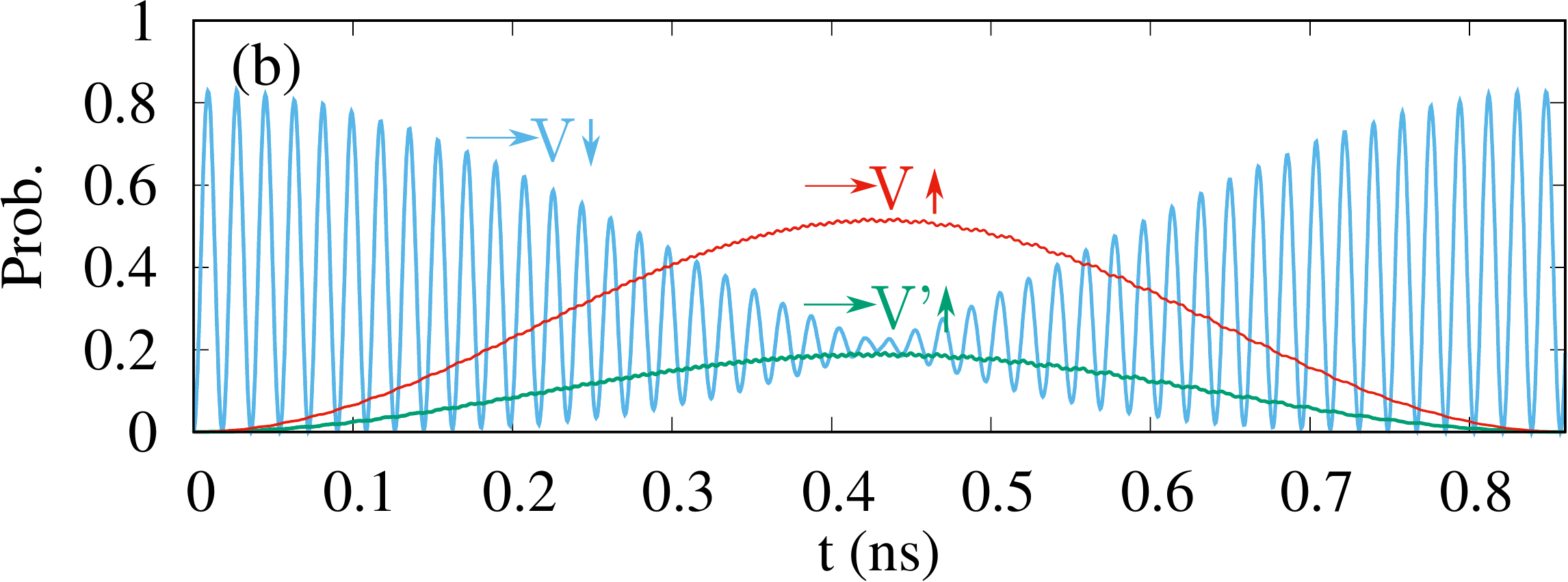}
 \caption{Solution of the electron dynamics for the monolayer graphene flake with a single excess electron in the conduction band driven
 by an external electric field.
The amplitudes are $F=0.8$ kV/cm (a) and $F=1.2$ kV/cm (b), the driving frequency is marked by an arrow in Fig. \ref{Trans2}(b) and (c).
$V'\downarrow$ ground state is the initial condition. 
 The lines show $|\langle \Psi(t)| \Psi_n \rangle|^2$ 
-- the square of the absolute value of the projection
of the time-dependent wave function $\Psi(t)$ on the lowest-energy conduction band eigenstates $\Psi_n(t)$ of the stationary Hamiltonian which are displayed in Fig. \ref{flakee}. 
The projection on the ground (initial state) $V'\downarrow$ is skipped. }
\label{czas} 
\end{figure}
\begin{figure}[hbt]
\vspace{1cm}
	\includegraphics[scale=0.28]{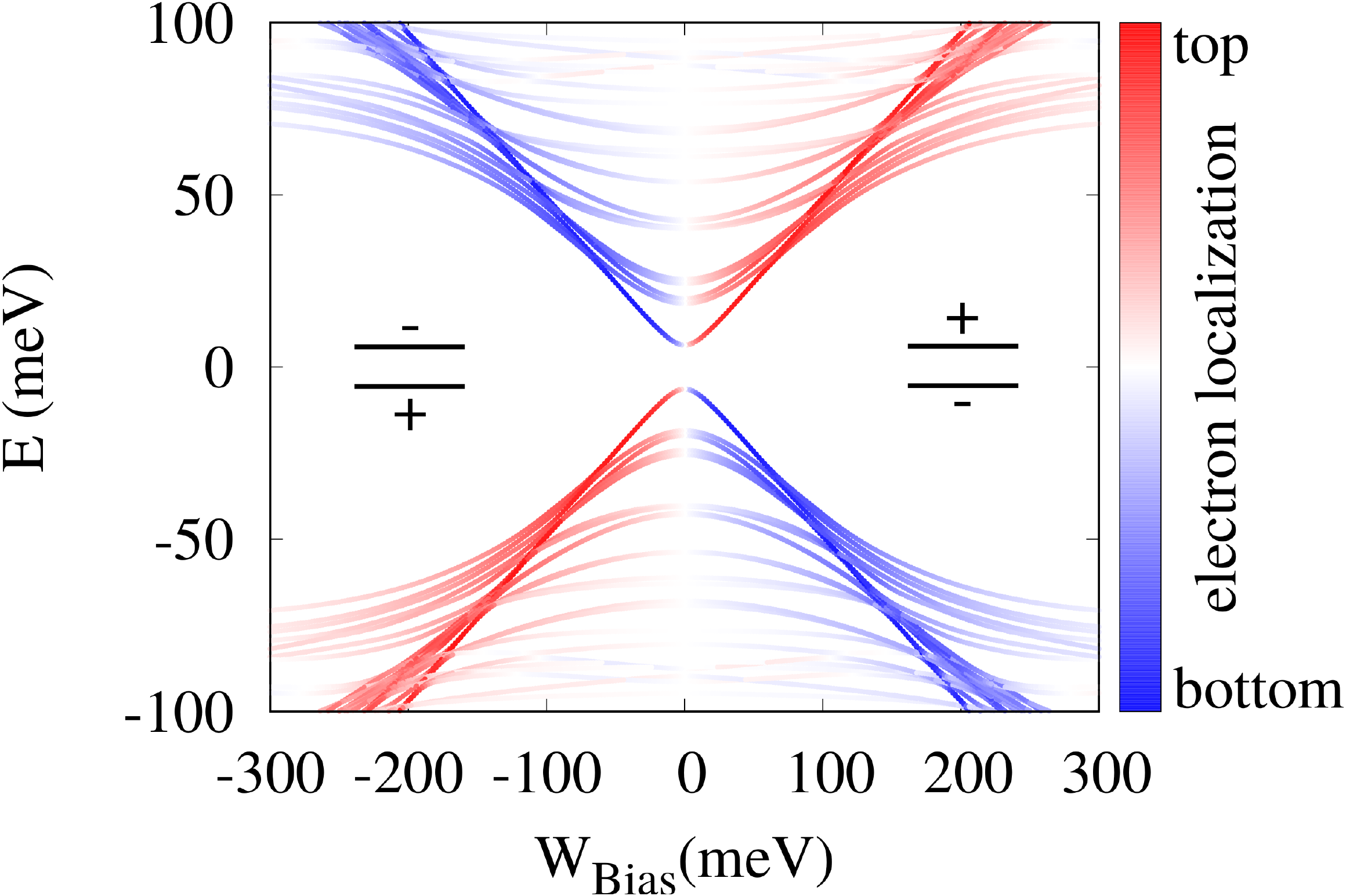}
 \caption{Energy levels of a hexagonal bilayer flake (side length 36.4 nm, 88212 carbon atoms within the flake) as a function of the bias between the layers. 
 No potential variation within each of the layers is introduced and no adatoms are present. The insets indicate the electron potential energy in the upper and lower layers
and the color of the lines the localization of the single-electron state in the bottom (blue) or red (top) layers respectively. }
\label{fbila}
\end{figure}
\begin{figure*}[hbt]
\begin{tabular}{ccc}
\includegraphics[scale=0.24]{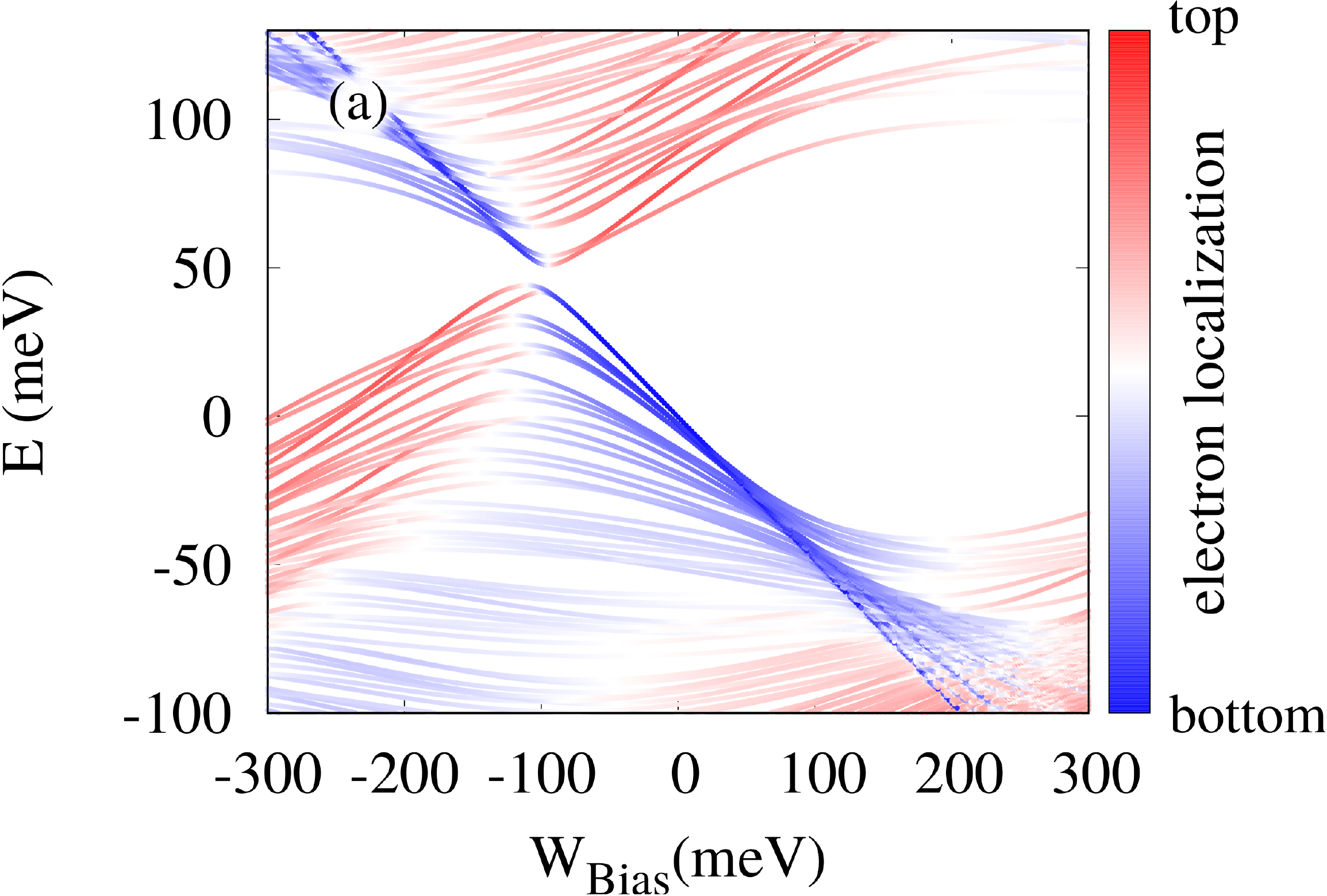}  \includegraphics[scale=0.24]{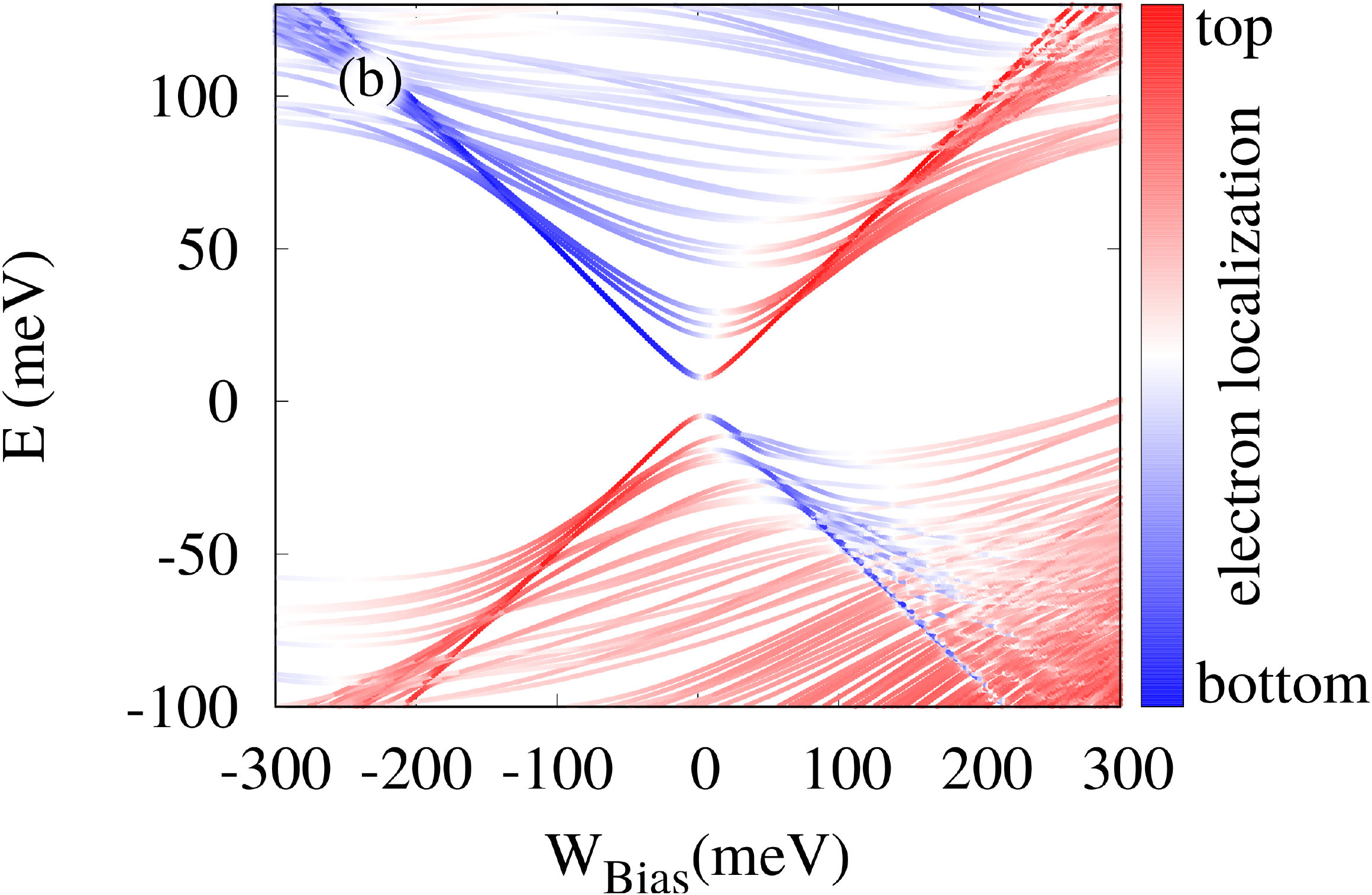} \includegraphics[scale=0.24]{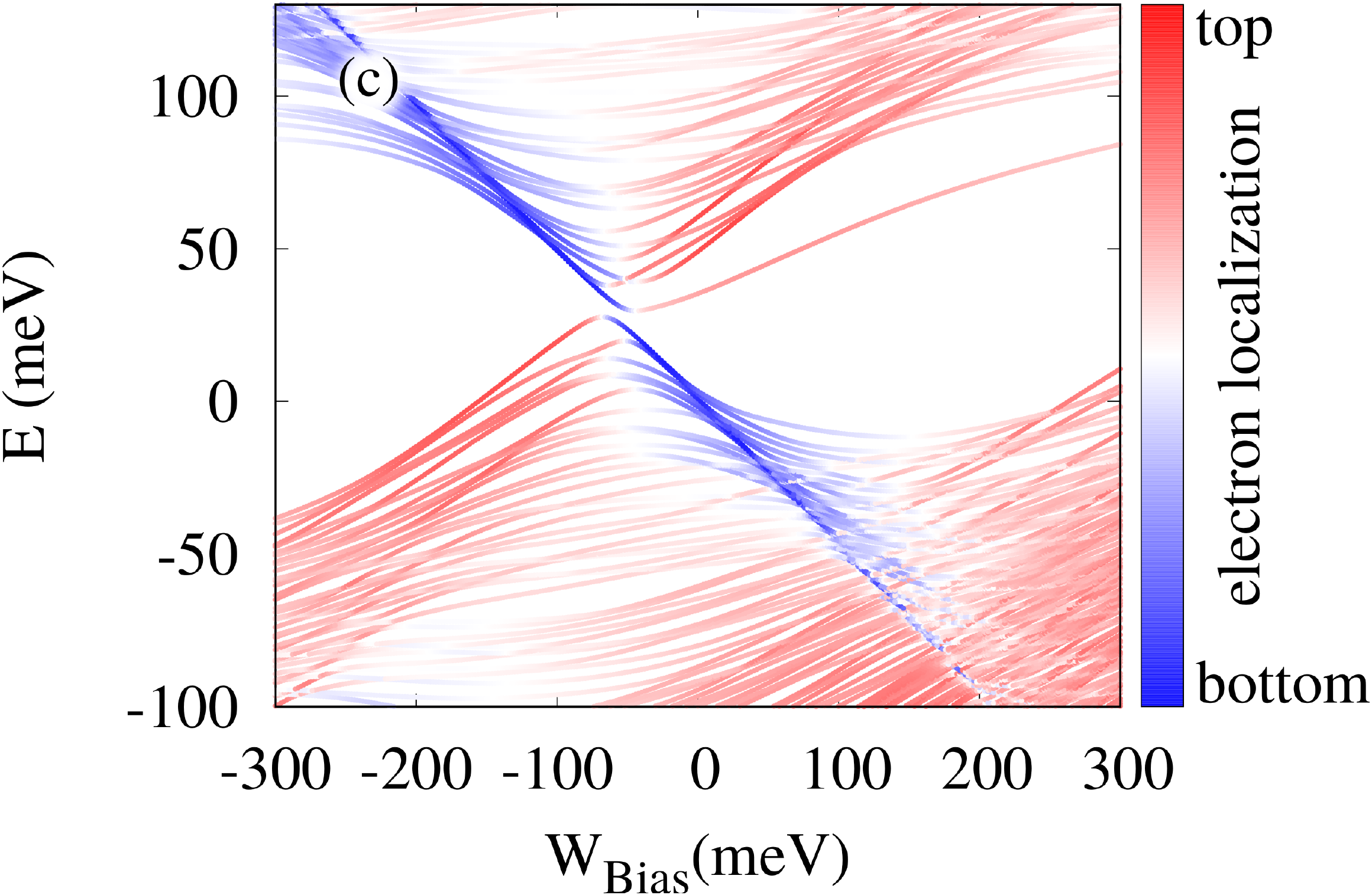} \\
\includegraphics[scale=0.24]{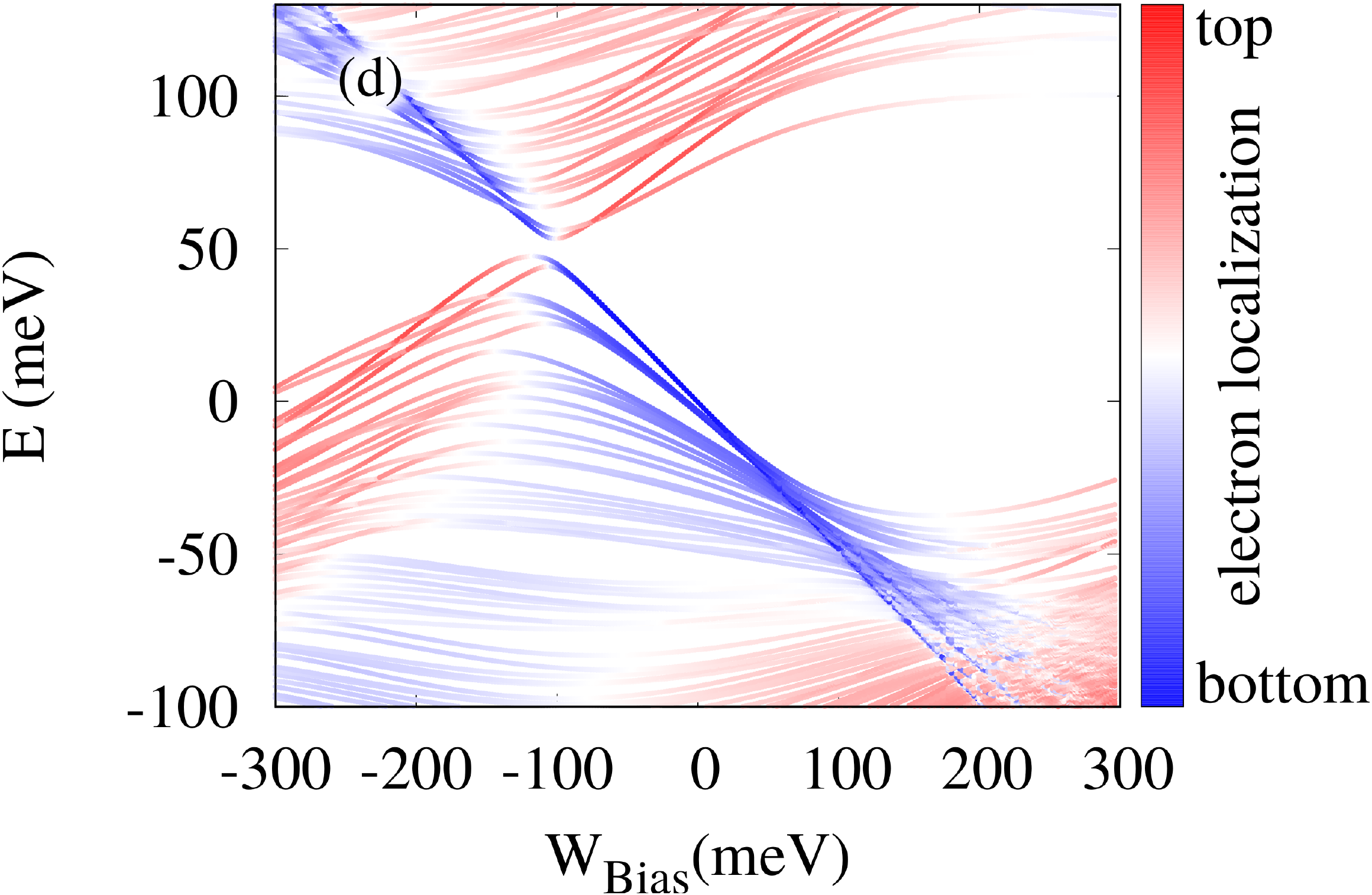}  \includegraphics[scale=0.24]{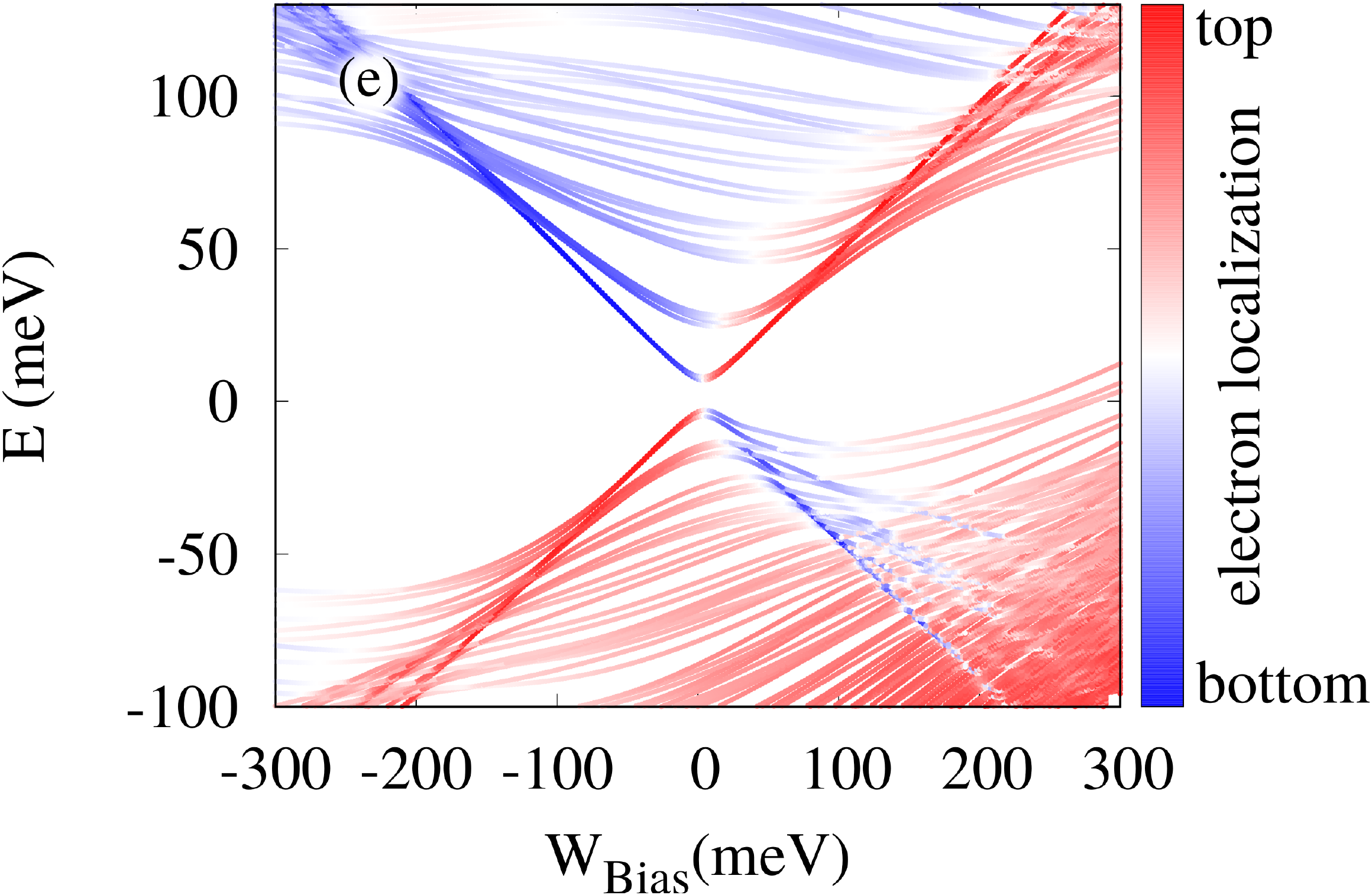} \includegraphics[scale=0.24]{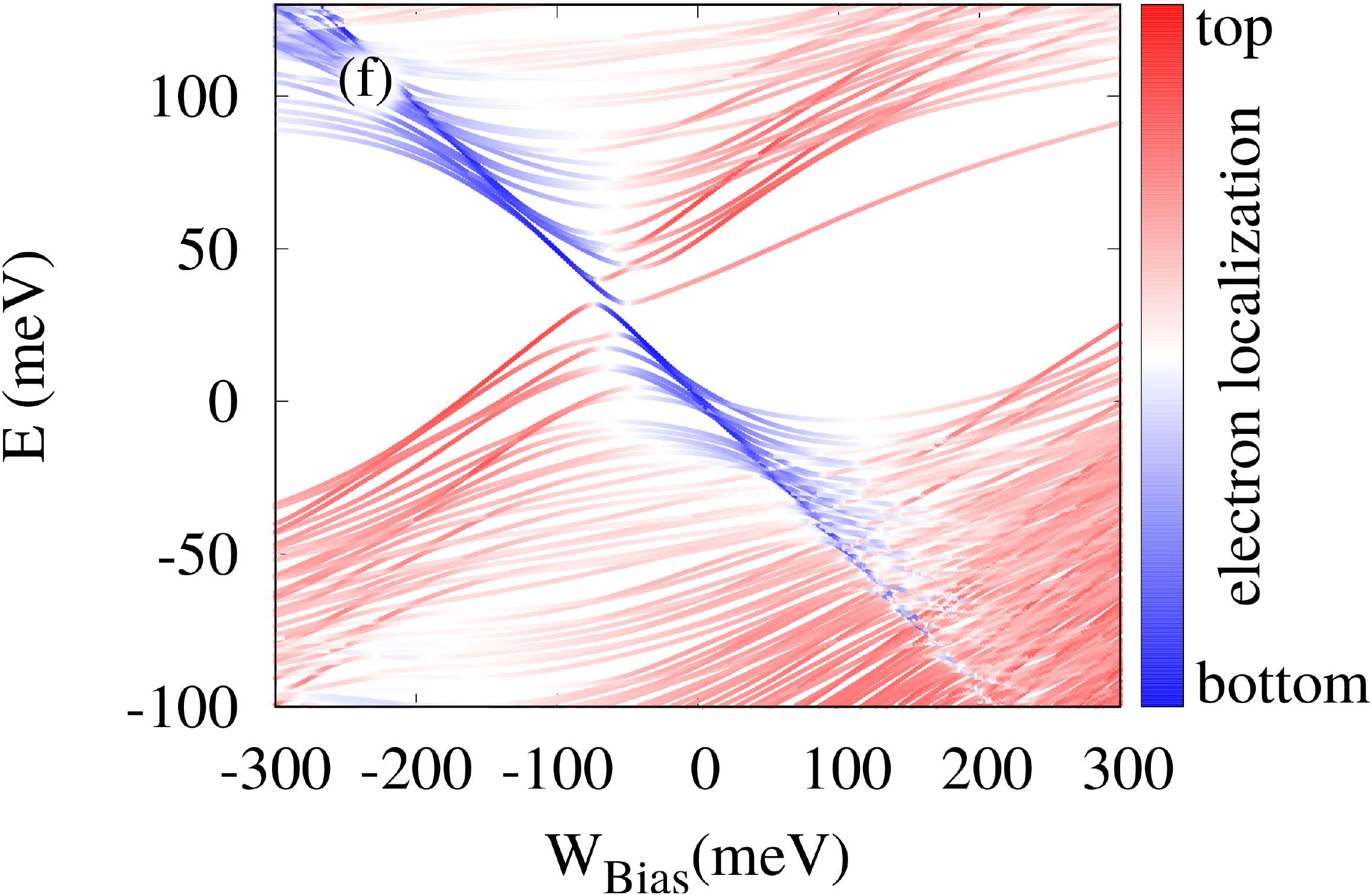} \\

\end{tabular}
\caption{Same as Fig. \ref{fbila} only with 221 fluorine atoms  adsorbed on the upper layer by (a,d) B2 atoms, (b,e) A2 atoms, and (c,f) at both B2 and A2 positions
at random with 50\% probability. In (a-c) only the direct A2-B1 hopping elements are included and no correction for the length of the bond is assumed. In (d-f) the skew
hopping elements and the Harrison correction to the bond length is introduced. The colors indicate the localization of the states
at the upper or lower graphene layers.
}\label{bila}
\end{figure*}
\section{Results} 
\subsection{Monolayer flake}
\subsubsection{Stationary states}
We first consider a graphene monolayer without any external electric potential. 
We take a hexagonal flake with armchair edges which guarantees  \cite{HexTriang}  an energy gap due to the finite size of the system. 
The considered hexagonal flake has a side length of about 10 nm  containing $\sim 12700$ carbon atoms and exhibits an energy  gap of about $\sim 150$ meV.
The energy levels near the gap are displayed in Fig. \ref{flake}(a) for $B_z=2$T. The energy levels marked with red (green) symbols were obtained for the graphene
flake without (with) the fluorine adatoms. We assume that 63 fluorine atoms are deposited at random positions within the flake, e.g. 0.5\% of carbon atoms are bound to the fluorine [Fig. \ref{flake}(d)]. 
We study the transitions within the quadruple of the lowest-energy levels of the conduction band, which appear between the ones marked with "b" and "c" in Fig. \ref{flake}(a) -- for the fluorinated
graphene flake and between "e" and "f" for the clean flake. 
The energy spectrum as a function of the perpendicular magnetic field is displayed in Fig. \ref{flakee}. 
The lowest energy levels of the conduction band at $B=0$ in the absence of the adatoms are four-fold degenerate with respect to the spin and the valley (see the red  lines in Fig. \ref{flakee}). 
The perpendicular magnetic field splits this fourfold degeneracy -- into nearly degenerate pairs of energy levels. The small splittings within each of the pairs are due to the spin Zeeman effect,
and the large splitting results from the interaction of the magnetic dipole generated by the electron currents [Figs. \ref{flake}(b,c)]. 
In the absence of the fluorine adatoms the armchair termination of the flake mixes the $K$ and $K'$ valleys \cite{HexTriang} but does not lift the valley degeneracy of the lowest quadruple of energy levels at $B=0$ (see Fig. \ref{flakee}). 
We will refer to the valley degree of freedom of the states that go down and up on the energy scale with $B_z$ as $V$ and $V'$ states, respectively.

The fluorine adatoms introduce a short range potential that is quite strong  -- see the
shift of the spectrum in Fig. \ref{flake}(a). The adatom can be treated as a local potential minimum which attracts the electron charge -- in particular
to the ions that are adjacent to the fluorinated one (see Fig. \ref{edens}).
 The adatoms placed at random locations perturb the equivalence between the A and B sublattices 
-- see Fig. \ref{edens} --  and thus leads to a lifting of the valley degeneracy at $B=0$. The four-fold degeneracy split into two Kramers doublets -- see the green lines in Fig. \ref{flakee}. 
The current circulation within the flake is hampered by the perturbation introduced by the fluorine adatoms [Figs. \ref{flake}(d,e)].
This leads to a  reduction of the orbital magnetic moments ($\boldsymbol \mu$) and a smaller slope of the energy levels ($\Delta E=-\boldsymbol \mu \cdot \boldsymbol B$).

\subsubsection{Spin transitions in the monolayer flake}
We consider  transitions from the lowest-energy state to the remaining three states of the quadruple (Fig. \ref{flakee}) that are driven by an in-plane electric field. 
Only the fluorinated flake is considered for this purpose. The transitions within the quadruple involve either valley or spin transitions which are induced  only by the electric field  when adatoms are present.
We solve the time-dependent Schr\"odinger equation over a time span of 2 $\mu$s and look for the maximal projection of the wave function on the stationary eigenstates.
The results for the lowest amplitude of the electric field $F=0.2$ kV/cm are displayed in Fig. \ref{Trans2}(a) as a function of the driving frequency.
 We can see narrow peaks for the intravalley (green) and intervalley (red) spin flips. The blue peaks correspond to intervalley transitions with conserved spin state. 
The peak near 12 meV corresponds to the direct, i.e. a single-photon, transition and the one near 6 meV is the second-order, two-photon transition \cite{mf1,mf2,mf3,mf4,mf5}. 

As the amplitude is increased to $F=0.8$ kV/cm [Fig. \ref{Trans2}(b)]  and $F=1.2$ kV/cm [Fig. \ref{Trans2}(c)] we notice (i) the  appearence of the three-photon transition
to  $V\downarrow$ state, (ii) a two-photon transition to  $V\uparrow$ -- marked with $\frac{V\uparrow}{2}$, (iii) a reduction of the direct transition probability to $V\uparrow$ state below 1 and (iv) the appearance
of the off-resonance transition to  $V'\uparrow$ state near $\hbar \omega= 12$ meV. 
The features (i) and (ii) result directly from an increase of the AC perturbation, while (iii) and (iv) can be understood on the basis of the time-resolved dynamics
of the system which is displayed in Fig. \ref{czas} for the frequency set to resonant for the intervalley spin-flip from $V'\downarrow$ ground state to  $V\uparrow$ state. 
The reduction of the transition probability below 1  for $F\geq 0.8$ kV/cm results from the participation of all the states of the quadruple in the time dynamics for $\hbar \omega=12$ meV.
Under these conditions the transitions have no longer a two-level Rabi character.
At $F=0.8$ kV/cm the intervalley spin-flip $(V'\downarrow)\rightarrow (V\uparrow)$ transition appears within the much wider and faster [Fig. \ref{czas}(b)] spin-conserving intervalley transition  $(V'\downarrow)\rightarrow (V\downarrow)$ [see Fig. \ref{Trans2}(b)]. 
Thus, the wave function of the driven system contains not only the AC frequency-targeted $V\uparrow$ component [red curve in Fig. \ref{czas}(b)] but also a contribution of the  $V\downarrow$ state
which appears soon after the AC driving was introduced. Moreover, 
the $\hbar \omega$ energy is close to $(V\uparrow)\rightarrow(V'\uparrow)$ energy difference [see Fig. \ref{flakee}]. This transition -- a spin conserving intervalley transition -- has a large rate
and width on the $\omega$ scale. After the transition from the $V'\downarrow$ ground state to $V\uparrow$, and the mediation of the latter,  the $V'\uparrow$ state appears in the dynamics. 
A closer inspection of the spin-flip transitions near $\hbar\omega=12$ meV  [see the inset to Fig. \ref{czas}(b)] we notice a smaller peak
for driving energy near $\hbar\omega=11.7$ meV - for which the spin flip transitions from $V'\downarrow$ to both $V\uparrow$ and $V'\uparrow$ states coincide.
Both the final states $V\uparrow$ and $V'\uparrow$ have the same spin orientation. 
The paths leading to the spin-flip transitions are different. The one for the higher energy peak was described in the preceeding paragraph.
For the lower energy peak the path for the transition goes via the spin-conserving and wide  $(V'\downarrow)\rightarrow V\downarrow$ transition,
then to a resonant intervalley spin flip  $(V\downarrow)\rightarrow (V'\uparrow)$, which requires a lower resonant frequency then the direct spin-flip $(V'\downarrow)\rightarrow (V\uparrow)$.
The state $(V'\uparrow)$ is then strongly coupled to $V\uparrow$, so they appear simultaneously in the dynamics. 
For  $F=1.2$ kV/cm [Fig. \ref{czas}(b)] the intervalley spin-conserving transition to the $V\downarrow$ state starts to dominate  the spin-flip transitions. For $F=2$ kV/cm 
the spin flips near $\hbar\omega=12$ meV [Fig. \ref{Trans2}(d)] are quenched. Then, the intervalley spin-flip transitions can only occur via the two-photon process
near half the nominal resonance energy [see the red peak near 6 meV in Fig. \ref{Trans2}(d)].
We find that the transition period, defined by the time for which the occupation of the targeted final state
is maximal -- even if below 100\% -- is inversely proportional to the driving amplitude $F$ as for the Rabi transitions. 
For $F=0.2$ kV/cm and $F=1.2$ kV/cm the spin-flip transition times from the $V'\downarrow$ ground state: 
1) to $V'\uparrow$ are: 256 ns, and 42.9 ns;
and 2)  to $V\downarrow$: 2.5 ns, and 420 ps. Since the spin relaxation times in monolayer graphene are of the order of 1.5 ns \cite{mass},  the latter spin-valley transition - seems within  experimental reach.
Note, that for $F=1.2$ kV/cm  the maximal transition probability is only 50\% [Fig. \ref{Trans2}(c)]. For larger amplitude of the AC electric fields the transition probability is strongly reduced [Fig. \ref{Trans2}(d)], and 
the two-photon transition is much slower than the direct one. 
For comparison the intervalley spin conserving transitions to $V\uparrow$ last 
 60 ps, and 10 ps, for $F=0.2$ kV/cm and $F=2$ kV/cm, respectively.
\subsubsection{Stationary states}
\subsubsection{Driven spin transitions}

For  bilayer graphene we consider a larger flake of side length 36.4 nm in order for the lateral confinement potential to fit within the structure. The number of carbon atoms is then 88212 with 221 fluorine atoms at the top layer. 

The energy spectrum for a biased bilayer flake is displayed in Fig. \ref{fbila}. The system of Fig. \ref{fbila} does not contain any adatoms
and a constant potential within the layers is assumed ($W_{QD}=0$). The energy gap found for zero bias is a finite size effect -- similar to the one found for the monolayer system in the preceeding section.
The bias applied between the layers enlarges this gap. The red and blue colors in Fig. \ref{fbila} indicate the electron localization at the top (bottom) layer. 

The spectrum for the fluorine adsorbed at the upper layer is given in Fig. \ref{bila}. 
In Figs. \ref{bila}(a-c) only the direct vertical (A2--B1) hopping elements -- were introduced. 
In Figs. \ref{bila}(d-f) the skew hoppings between A1 and B2 atoms were added and the corrections to the bond lengths due to the deformation of the fluorinated layer are accounted for. 
The energy effects of the correcting terms are small, but an influence is detected in the transition times -- particularly for the spin conserving transitions (see below).


The fluorine atoms when deposited on the B2 sites only [cf. Figs. \ref{bila}(a,d)] -- i.e. 
at atoms that do not form  dimers with the bottom layer -- shift the position of the minimal band gap to 
negative biases. The fluorine over the B2 atom attracts the electrons to its A2 neighbors - which as bound to the bottom layer drain the electron charge from below. A negative bias -- that 
shifts up the potential at the top layer with respect to the lower one needs to be applied to restore the symmetry between the layers, hence the shift of the minimal energy gap. 
On the other hand, when  fluorine is adsorbed over the A2 sites -- the enhancement of the electron charge at A2 is found only a slight [see Fig. \ref{edens}]. The charge is increased and its neighbors -- B2 atoms -- do not form  dimers
with the bottom layer, hence the minimal band gap stays at zero bias [cf. Figs. \ref{bila}(b,e)]. 
The binding energy of the fluorine atom to the B2 in the dilute limit was found \cite{Santos} only slightly larger than the binding to A2 atoms. The difference in the energy is small
and decreases when the concentration of fluorine is reduced  (0.4 meV for a fluorium atom per 18 carbon atoms only \cite{Santos}). Hence,
an equal distribution of the  fluorine adatoms on both A2 and B2 sublattices seems most likely. The energy spectrum for this case [cf. Fig. \ref{bila}(c,f)] resembles an average of the spectra for
adatoms bound exclusively  to A2  [cf. Figs. \ref{bila}(b,e)]  and B2  [cf. Figs. \ref{bila}(a,c)]  and
exhibits the minimal band gap for a bias closer to zero than for the fluorine adsorbed on the B2 sublattice. 

\begin{figure}[hbt]
\hspace{-0.5cm}\includegraphics[scale=0.35]{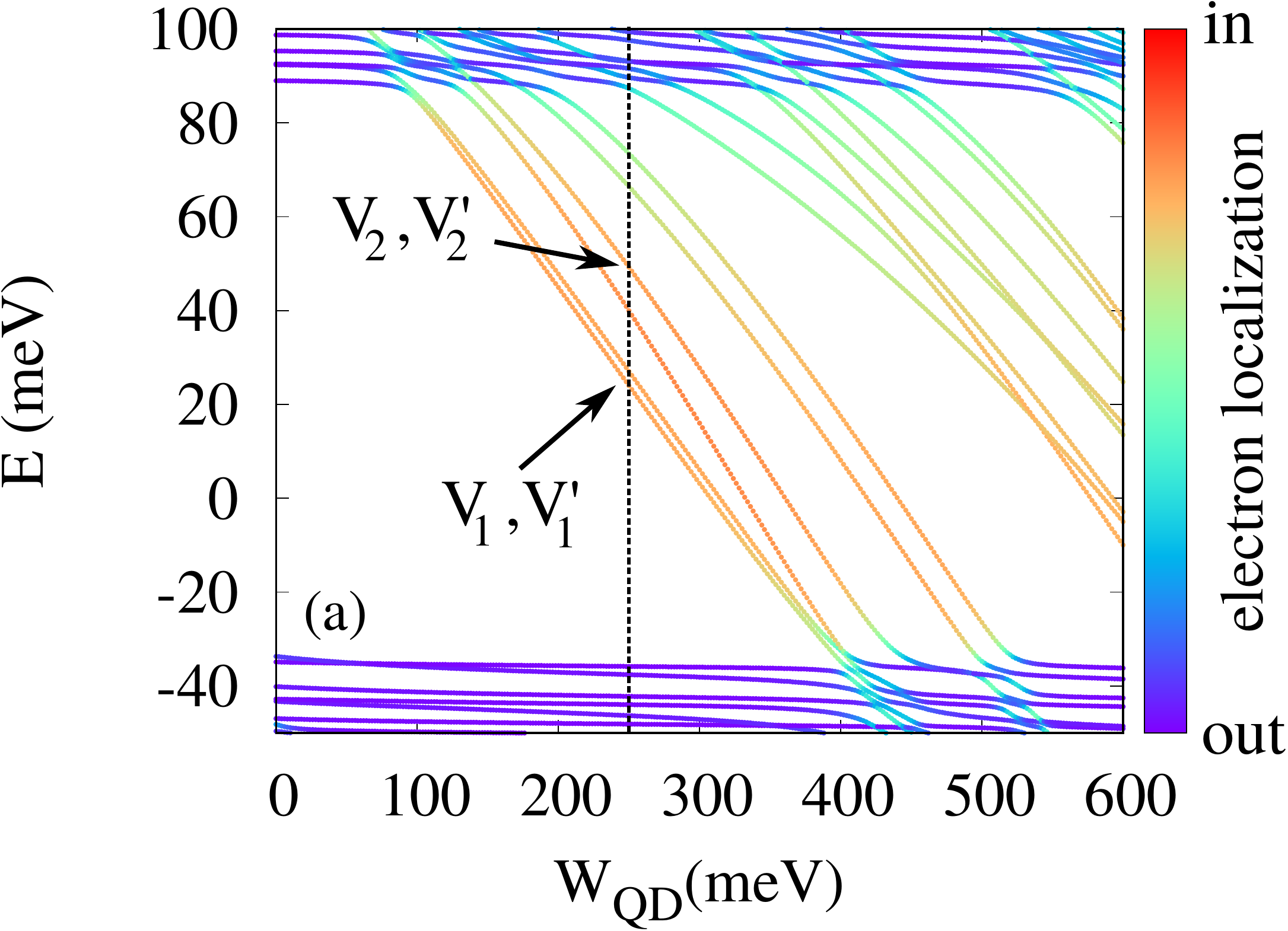} \\
\hspace{-0.5cm} 	\includegraphics[scale=0.35]{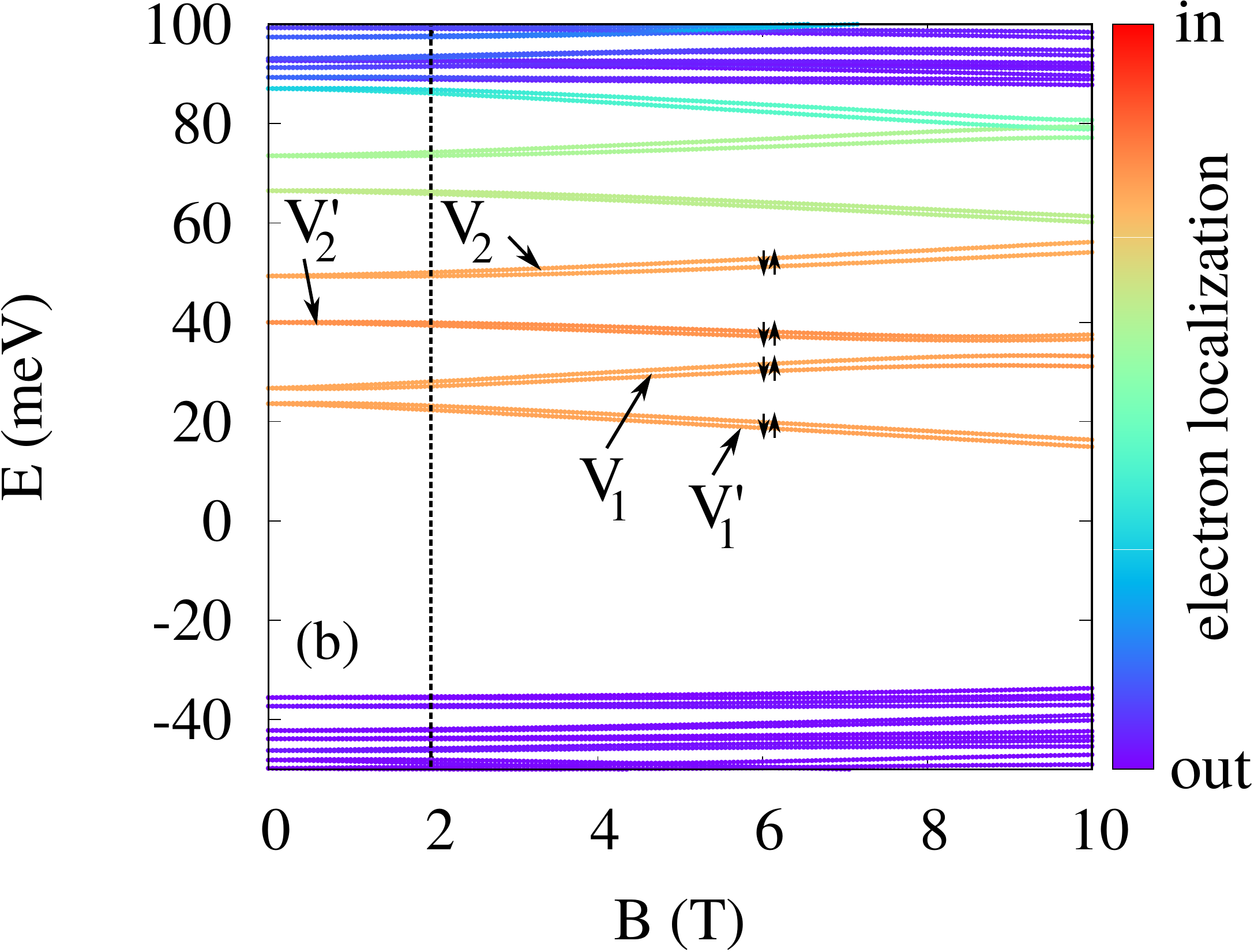} \\
\hspace{-0.5cm} \includegraphics[scale=0.35]{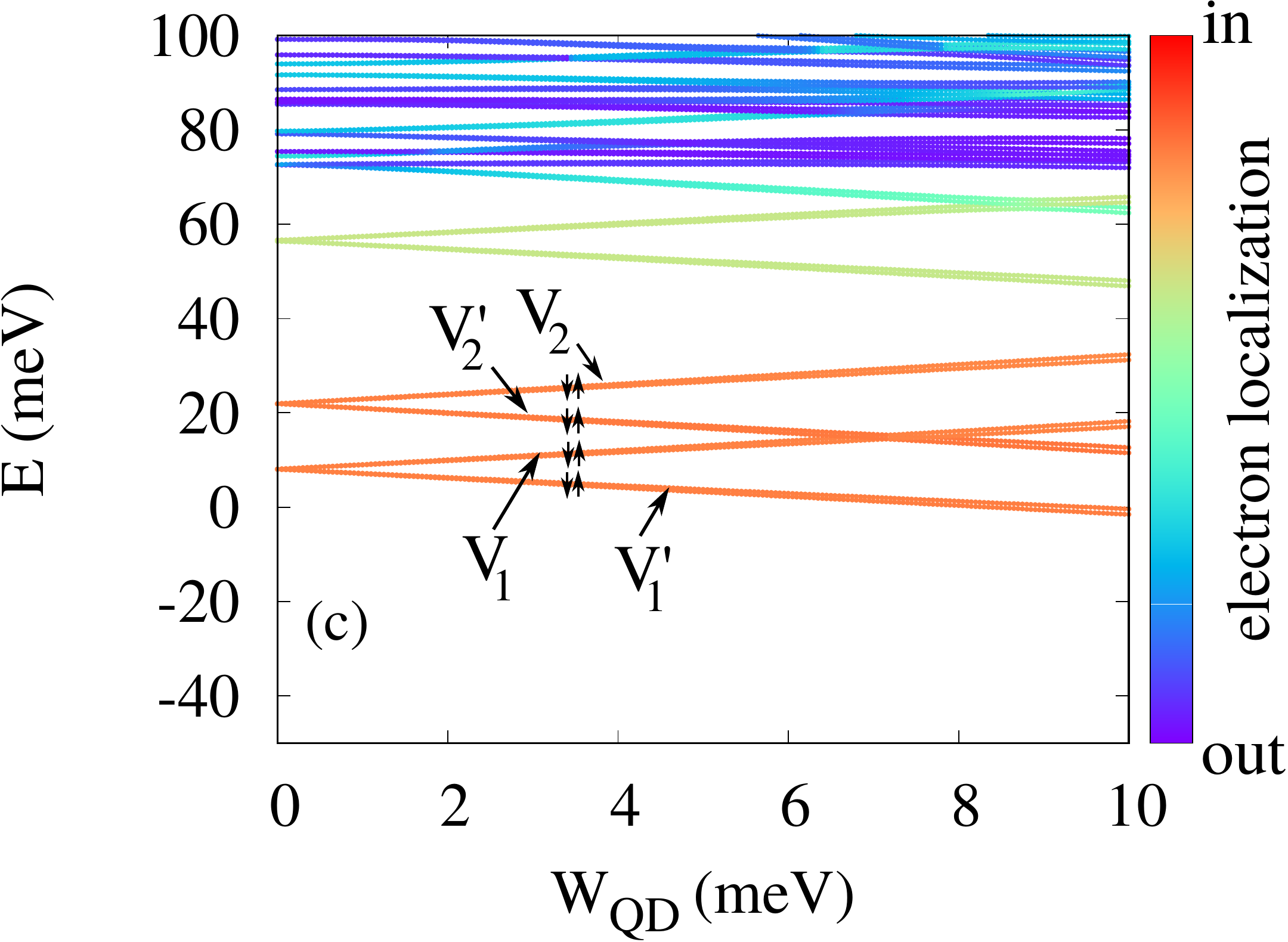} \\
 \caption{(a) The energy spectrum as a function of the lateral potential $W_{QD}$ for the bias between the layers $W_{bias}=-300$
 meV
in the presence of the fluorine adatoms. (b) The lowest-energy quantum dot-confined energy levels in perpendicular magnetic field. (c) Same as (b) only for a clean bilayer. 
The color indicates the electron localization inside and outside the quantum dot.
The red one indicates that the entire electron density is localized in the area closer to the QD center than $1.5r_0$.}
 \label{pp}
\end{figure}

Now, we are ready to introduce a lateral confinement potential to create the quantum dot. We will focus on the workpoint given by the bias voltage  $W_{bias}=-300$ meV.
As the lateral potential is introduced the QD confined states appear in the energy gap open by the bias -- see Fig. \ref{pp}(a).

The energy levels in Fig. \ref{pp}(a) are two-fold degenerate with respect to the spin. 
For the rest of the results we assume $W_{QD}=250$ meV, for which the lowest-energy states are in the center of the energy gap.
The energy spectrum as a function of the perpendicular magnetic field is displayed in Fig. \ref{pp}(b). The valley mixing effects
induced by the adatoms can be estimated by comparison with Fig. \ref{pp}(c), in which the adatoms were absent: the states at $B=0$ are four-fold degenerate and a crossing of energy levels instead of an avoided crossing occurs near 7T.

\begin{figure}[hbt]
\hspace{-0.5cm}\includegraphics[scale=0.35]{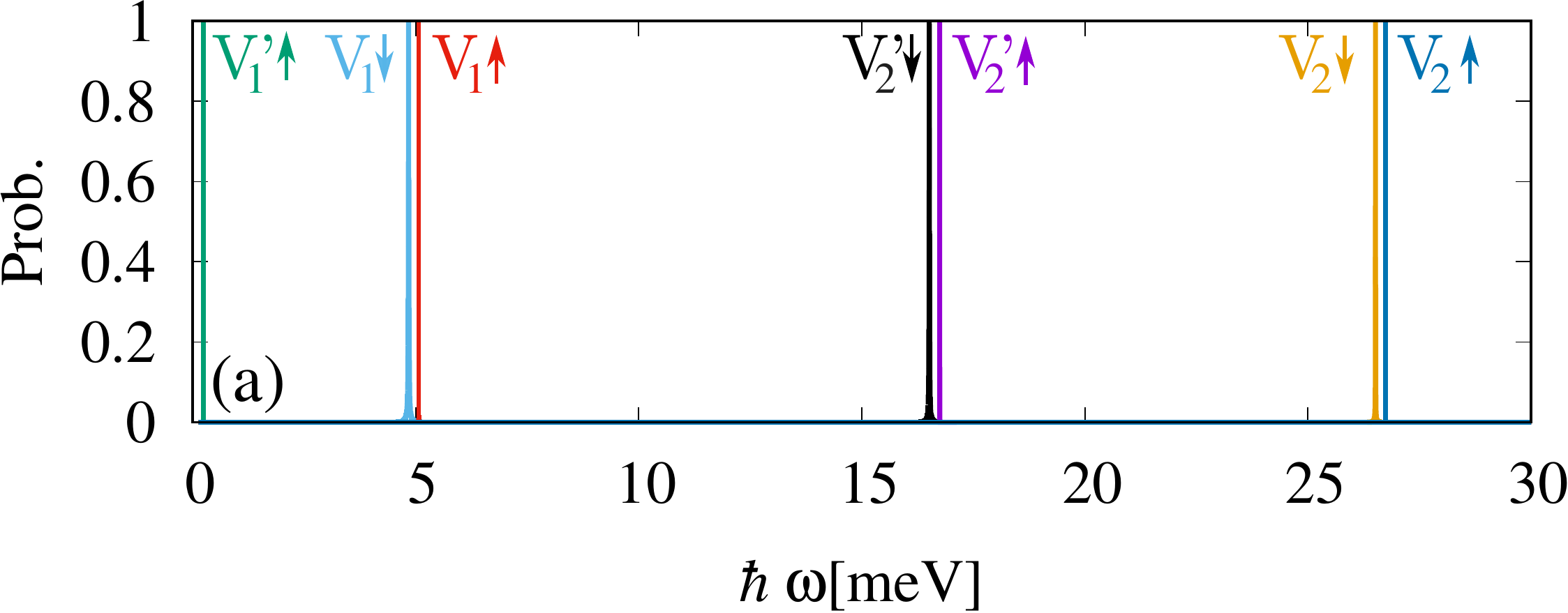} \\
\hspace{-0.5cm} 	\includegraphics[scale=0.35]{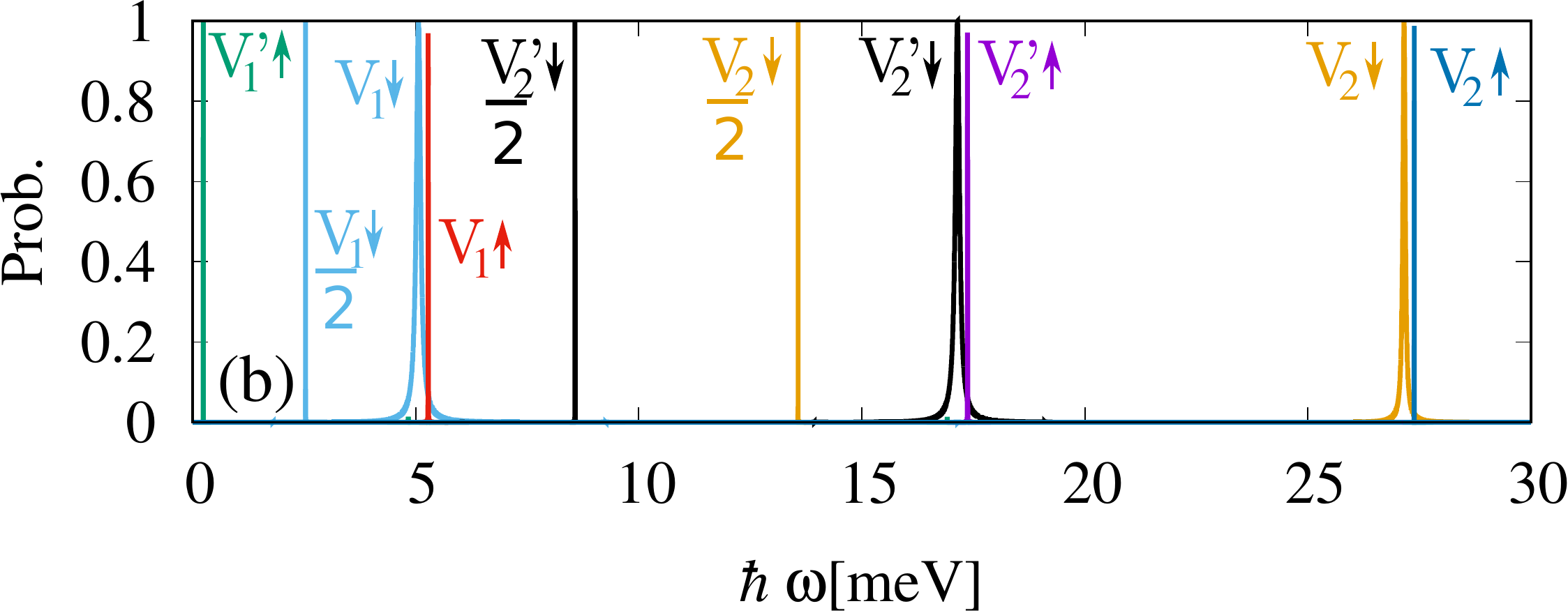} \\
\hspace{-0.5cm} \includegraphics[scale=0.35]{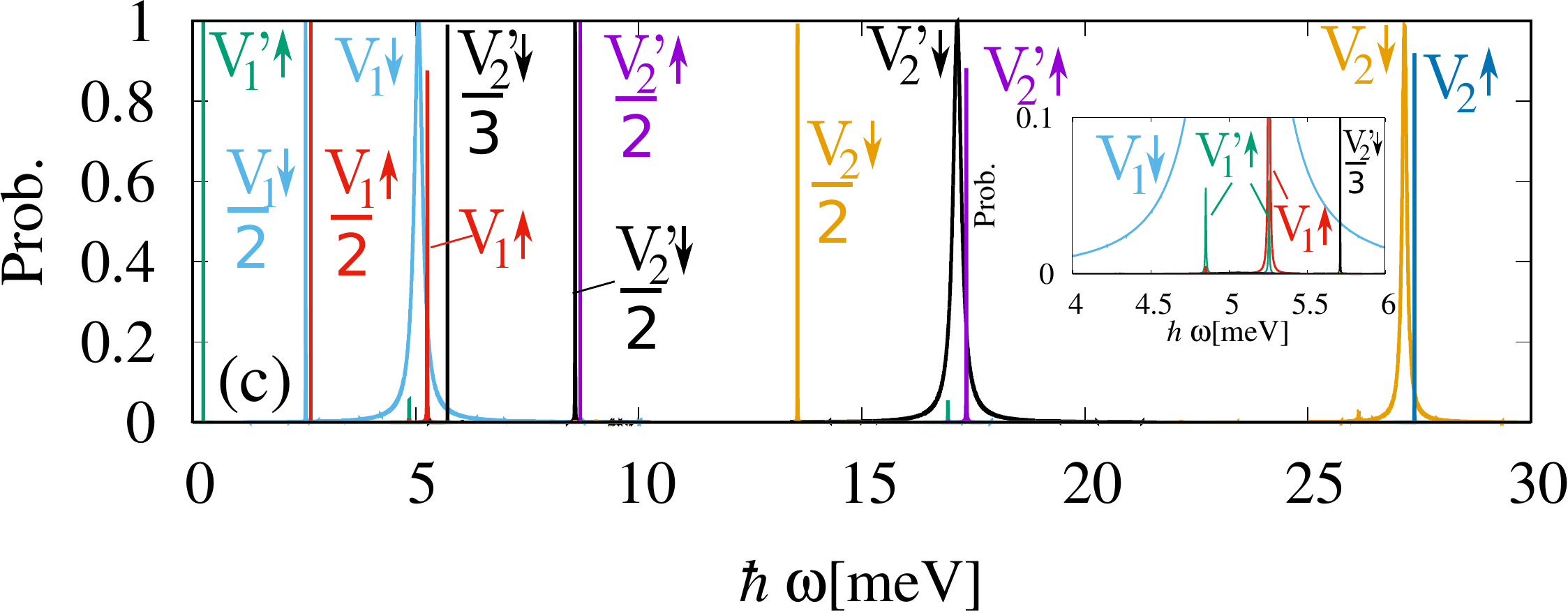} \\
 \caption{ Maximal occupation probabilities for the system in the perpendicular magnetic field of 2 T and  AC electric field  amplitude of $F=0.2$ kV/ nm (a), $F=2$ kV/nm (b), and $F=4$ kV/nm (c). For the initial condition the $V_1'\downarrow$ ground state is adopted. The simulation lasted for 2$\mu$s. } \label{pt}
\end{figure}

We  consider transitions between the energy levels marked by Fig. \ref{pp}(b) with the $V_1'\downarrow$ ground state as the initial condition. 
The maximal transition probability for the driving time of 2$\mu$s is given in Fig. \ref{pt} for the amplitude of the AC field $F=0.2$ kV/cm Fig. \ref{pt}(a), $F=2$ kV/cm [Fig. \ref{pt}(b)],  and $F=4$ kV/cm [Fig. \ref{pt}(c)]. 
For $F=0.2$ kV/cm only the direct Rabi transitions are observed [Fig. \ref{pt}(a)] to all the 7 energy levels that cover this energy range. 
For  $F=2$ kV/cm [Fig. \ref{pt}(b)] we notice  an increased width of the direct spin conserving transitions to $V_1\downarrow$, $V_2'\downarrow$ and $V_2\downarrow$. 
Moreover, the two-photon transitions at half the direct transition energy are obseved [Fig. \ref{pt}] to these three states. 
One also notices a slight reduction of the spin-flipping transition probabilities below 1 for $V_1\uparrow$, $V_2'\uparrow$ and $V_2\uparrow$ - that appear within the spin-conserving transition peaks. 
A similar effect was discussed above for monolayer graphene.  The maximal transition probability to $V'\uparrow$ and the low-energy limit is unaffected. 
For $F=4$ kV/cm, in addition to the preceeding results the three photon transition to $V_2'\downarrow$ is observed and the two-photon spin-flip transitions to $V_1\uparrow$ and $V_2'\uparrow$. 
Within the $V_1\downarrow$ and $V_2'\downarrow$ peaks a small $V_1'\uparrow$ 
peaks appear in the mechanism described above for the monolayer flake.


The transiton times within the range of amplitudes of Fig. \ref{pt} vary linearly with $1/F$ 
-- see Table \ref{tabu}. The values of $t$ in the Table correspond to only the vertical interlayer hoppings
included in the Hamiltonian with hopping parameters independent of the distance from a fluorine adatom.
The values of $t'$ account for the skew hoppings and the deformation of the top graphene layer by the adatom. 
In most cases -- with a few exceptions -- the transition times change only within 10\% when these
two corrections are introduced to the Hamiltonian. 
The spin-flip times $t'$ for $F=2$ kV/cm with the final state $V_2'\uparrow$ and $V_1\uparrow$ 
take 1.2 ns and 1.7 ns, respectively, and for $F=4$ kV/cm these times are
 0.6 ns and 0.84 ns, respectively. Note, that for $F=4$ kV/cm - the direct transition
probabilities to these two states are above 80\% [Fig. \ref{pt}(c)].  For the bilayer
flake the experimental relaxation times reach 3 ns \cite{weihan}, hence the ones obtained  for $F=4$ kV/cm 
gives  spin transition times that are by a factor of 5 to 6 shorter

\begin{table}
\begin{tabular}{c |c| c| c} 
final state & F [kV/cm] & $t$  & $t'$  \\ \hline\hline
$V_1'\uparrow $& 0.2 & 1.5 $\mu$s & 640 ns\\
$V_1\uparrow $& 0.2 & 16.5 ns & 12.5 ns\\
$V_2' \uparrow$ & 0.2 & 18 ns  & 17.4 ns\\
$V_2 \uparrow$ & 0.2 & 133 ns & 143 ns\\
\hline
$V_1'\uparrow $& 2 & 144 ns & 64.5 ns\\
$V_1\uparrow $& 2 & 1.6 ns & 1.2 ns\\
$V_2' \uparrow$ & 2 & 1.8 ns  & 1.7 ns\\
$V_2 \uparrow$ & 2 & 13.3 ns & 1.4 ns\\
\hline
\hline
$V_1\downarrow $& 0.2 & 364 ps & 293 ps\\
$V_2' \downarrow$ & 0.2 & 370 ps & 363 ps\\
$V_2 \downarrow $ & 0.2 & 652 ps & 886 ps\\
\hline 
$V_1\downarrow $& 2 & 36 ps & 30 ps\\
$V_2' \downarrow$ & 2 & 37 ps & 36 ps\\
$V_2 \downarrow $ & 2 & 66 ps & 88 ps\\
\end{tabular}
\caption{Transition times for the bilayer graphene quantum dot for the parameters considered in Fig. \ref{pt}(a,b) with the perpendicular magnetic field $B=2$ T, the intelayer bias of $W_{bias}=-300$ meV and the lateral confinement potential of depth $W_{QD}=250$ meV.
$V_1'\downarrow$ as the initial state and the amplitudes of the AC in-plane electric field $F=0.2$ kV/cm and $F=2$ kV/cm. The results in the third column $t$ were obtained with the Hamiltonian that contains only the direct vertical $A2-B1$ hoppings. In the last column $t'$ the Hamiltonian contains also the skew hoppings and the Harrison correction to the interatomic distances due to the defromation of the top graphene layer.} \label{tabu}
\end{table}

\section{Summary and conclusions}
We have investigated spin transitions driven by EDSR resonance in finite fluorinated flakes of monolayer graphene 
and  bilayer graphene quantum dots with electrostatic lateral confinement. 
The study was based on the atomistic tight binding approach and a direct numerical solution of the time-dependent Schr\"odinger equation.  
We  described the effects of the fluorine adatoms on the confined energy levels. 
The adatoms reduce the orbital magnetic dipole moments due to current circulation within the flake. 
Moreover, the adatoms induce strong intervalley scattering effects lifting the fourfold degeneracy of the confined ground states. 
The spin-transitions that are allowed by the strong local spin-orbit coupling interactions in the neighborhood of the adatoms even for the dilute concentration of the fluorine  reach sub-nanosecond range below the spin relaxation times in graphene, and which should allow for their observation in the  lifting of spin Pauli blockade of the current. 
The spin-flipping transitions compete with the spin-conserving ones with transition energies that differ only
by the spin Zeeman interaction energy. As a result of this competition the maximal spin-flip probabilities fall below 1
for larger amplitudes of the AC field, for which the times for the maximal spin flip probability remain
inversely proportional to the amplitude. When the direct spin-flip transitions are attenuated by their spin-conserving counterparts the spin-flip can be accomplished by  two-photon transitions.
 

\section*{Acknowledgments}
This work was supported by the National Science Centre
according to decision DEC-2013/11/B/ST3/03837 and by the Flemish Science Foundation (FWO-VL).

\bibliographystyle{unsrt}

\end{document}